\documentclass[sigconf]{acmart}

\usepackage{balance}

\usepackage{listings}
\usepackage{cleveref}
\usepackage{booktabs} 
\usepackage{xspace}
\usepackage{xcolor}
\usepackage{microtype}
\usepackage{subcaption}
\usepackage{balance}

\definecolor{darkgreen}{rgb}{0.15,0.55,0.15}
\definecolor{mred}{rgb}{.80,.12,.30}
\definecolor{grey}{rgb}{0.5,0.5,0.5}
\definecolor{Purple}{rgb}{.75,0,.85}
\definecolor{light-gray}{gray}{0.95}
\definecolor{mid-gray}{gray}{0.85}
\definecolor{darkred}{rgb}{0.7,0.25,0.25}
\newcommand{\red}[1]{\textcolor{red}{#1}}
\newcommand{\redtt}[1]{\textcolor{red}{\texttt{#1}}}

\newcommand{\eat}[1]{}

\newcommand{\ititle}[1]{\smallskip\noindent\emph{#1}}
\newcommand{\stitle}[1]{\smallskip\noindent\textbf{#1}}
\newcommand{\sstitle}[1]{\noindent\textbf{#1}}

\newlength{\listingindent}                
\usepackage[LGRgreek]{mathastext}

\newtheorem{defi}{Definition}

\newtheorem{example}[defi]{Example}

\setcopyright{none}

\makeatletter
\def\@copyrightspace{\relax}
\makeatother

\renewcommand\footnotetextcopyrightpermission[1]{} 
\begin{document}
\title{Provenance for Interactive Visualizations}

\eat{\titlenote{A version of this paper has been accepted to the 3rd Workshop on Human-In-the-Loop Data Analytics (HILDA '18).}}

\author{Fotis Psallidas}
\affiliation{%
  \institution{Computer Science, Columbia University}
}
\email{fotis@cs.columbia.edu}

\author{Eugene Wu}
\affiliation{%
  \institution{Computer Science, Columbia University}
}
\email{ewu@cs.columbia.edu}

\begin{abstract}
We highlight the connections between data provenance and interactive visualizations.  To do so, we first incrementally add interactions to a visualization and show how these interactions are readily expressible in terms of provenance.  We then describe how an interactive visualization system that natively supports provenance can be easily extended with novel interactions.
\end{abstract}

\maketitle

\section{Introduction}
\label{s:intro}
Interactive data visualizations enable users to rapidly recognize important patterns within the data, by leveraging the powerful capabilities of the human perceptual system, and to identify and explore salient relationships that are not readily evident from a static visualization. As such, they constitute a cornerstone in many human-in-the-loop data analysis and management systems across domains including data exploration and decision-support~\cite{endeca,powerbi}, knowledge exploration~\cite{yagobrowswer2011tylenda,althoff2015timeline}, debugging and analysis of machine learning and statistical models~\cite{tensorboard,shiny,seq2seqvisv1}, interactive data cleaning~\cite{kandel2011wrangler,kandel2012profiler,scorpion,wu2012demonstration} and profiling~\cite{nadeef:2013:ebaid:2013,metanome:2015:papenbrock}, to name a few.

The increasing importance and ubiquity of interactive visualization tools, along with the massively increasing scale of modern datasets, has seen a convergence between the visualization and database communities.   Visualization systems incorporate data processing capabilities such as filtering, grouping, aggregation, ordering, and scaling in order to compute data summaries that are further rendered on the screen.  However, increasing dataset sizes has caused data processing to become a core bottleneck that impedes interaction responsiveness and is detrimental to the overall data analysis and exploration of end-users~\cite{shneiderman:1984,hanrahan2012adt,heer2012idv}.

\begin{figure}[t]
\centering
 \includegraphics[width=\columnwidth]{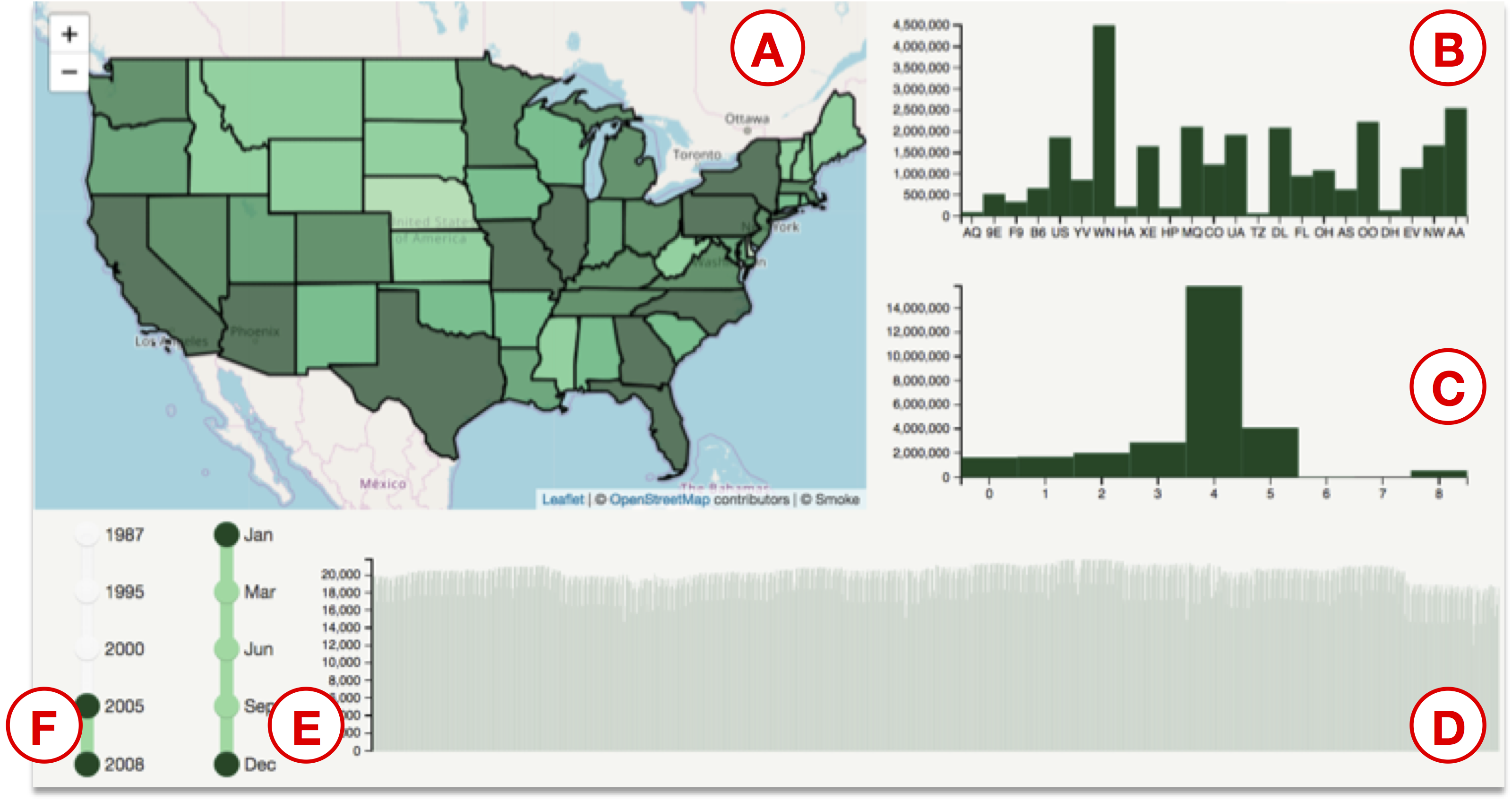}

\caption{\small Example of an interactive visualization.}
 \label{fig:intro}

\end{figure}

To this end, recent work in both communities has proposed systems to combine query processing and visualization functionality within a single framework.   For instance, Reactive Vega~\cite{reactivevega} draws upon stream query processing and declarative languages (i.e., Vega-lite~\cite{satyanarayan2017vegalite}) to model the data processing, visualization, and interaction processes within a unified dataflow framework.  Similarly, the Data Visualization Management System~\cite{EugeneWuVision2014,wu2017dvms} proposes a relational abstraction to model interactive visualizations as relational workflows that map database relations and relations of user events to marks and, ultimately, pixels on the screen.   For instance,  consider the multi-view interactive visualization of~\Cref{fig:intro}:

\begin{example}[Exploring Flight Delays]\Cref{fig:intro} visualizes a breakdown of delayed flights~\cite{ontime} coupled with a crossfilter interaction technique~\cite{crossfilter}. Each chart renders the output of a count aggregation query of delayed flights grouped by different attributes: by state (A), airline (B), departure delay (C), date (D), month (E), and year (F).  Thus, the visualization may be modeled as a large relational workflow composed of these aggregations, along with visualization workflows to map the results to visual marks (e.g., rectangles, circles, or polygons) which, in turn, are mapped to pixels on the screen. Crossfilter interactions let users select data in any of the views and see the other views update to show the statistics represented by the selected subsets.
\end{example}

Drawing the connection between relational workflow processing and interactive visualization not only improves the productivity of developers by introducing higher level languages to express visualizations, but has led to a rich area of performance-oriented research. Recent research efforts adapt query optimization techniques to the visualization domain and develop novel techniques inspired by unique characteristics of visualizations.  These include adapting columnar execution~\cite{kandel2012profiler}, perception- and visualization-aware online aggregation~\cite{procopioload,alabipfunk,kim2014rapid,rahman2017ve}, speculative exploration sampling~\cite{dice2014}, and visualization  prefetching~\cite{battle2016}, to name a few.   Notably, most of this work has been focused on speeding up specific visualization interactions or specific classes of database queries.

In this paper, we build on this convergence by highlighting the connection between data provenance~\cite{provdb} and visualization interactions.  Provenance broadly describes the process by which data artifacts are created and transformed.  In the context of a relational workflow, it both describes the sequence of operators that transformed input relations to result relations (coarse-grained provenance), as well as relationships between individual input and output records of the workflow (fine-grained provenance or lineage).

The use of provenance in visual analytics is not new. Previous efforts leveraged coarse-grained provenance of data, interactions, and visualizations in the form of histories that can be used to support collaborative communication, replication and reproducibility, action recovery, sense-making, and meta-analysis (see survey~\cite{ragan2016characterizing}  and tutorial~\cite{herschel:2016:provviz_tutorial}).  Unfortunately, the role of fine-grained provenance in interactive visualization has been less explored.  We believe that a major factor is performance: the overhead to track fine-grained provenance can slow fast query processing engines by multiple orders of magnitude and cripple interaction response times. 

To this end, recent work demonstrated a fine-grained provenance-enabled relational engine~\cite{psallidas2018smokedemo,psallidas2018smoke} that is fast enough, and incurs sufficiently low overhead, to out-perform specialized interactive visualization systems on cross-filtering benchmarks and maintain sub-$100ms$ interaction times on a 123.5M row flight dataset. These results illustrate the feasibility of expressing interactive visualizations using high-level provenance constructs, while also benefiting from fast execution engines.  Following this, the purpose of this paper is to explore two questions: \textbf{how can leveraging provenance concepts make it easier to build existing interactive visualizations?}, and \textbf{does taking a provenance perspective enable {\it new} interactions and visualization interfaces that are otherwise challenging to express?}

The rest of the paper is split into two sections.  Section~\ref{s:express} introduces the connections between interactive visualizations (and when possible, interactive applications in general) and provenance concepts.  To do so, we start with a trivial non-interactive visualization, and incrementally endow it with different types of interactions commonly found in the information visualization literature.  For each, we will describe how it is currently constructed, draw its connection with provenance, and remark on details regarding performance or semantics.     Section~\ref{s:crazy} builds upon this perspective by exploring how expressing and implementing interactive visualizations on top of provenance-enabled visualization engines can leverage existing provenance analysis research and greatly extend the expressive power of interactive visualizations.
\section{Interaction As Provenance}
\label{s:express}

Interactive visualizations can be modeled as workflows that map between the data and the pixel space.  User interactions can be viewed as dynamically transforming the workflows, or rapidly creating new workflows, and ultimately cause changes in the pixel space~\cite{wu2017dvms}. In this section, and along this conceptual model, we illustrate the connections between provenance---in particular, fine-grained provenance between individual input and output records---and visualization interactions. (We use provenance and fine-grained provenance interchangeably, and clearly state when we refer to coarse-grained or other forms of provenance semantics.) To better explain the connections, we progressively build interactive visualizations over the following database schema of delayed flights:

{\small
\begin{lstlisting}[
	language = SQL,
	showspaces=false,
	basicstyle=\ttfamily\footnotesize,
	commentstyle=\color{gray},
	mathescape=true,
	numbers=none,
	frame = single,
	escapeinside={<}{>},
	captionpos=b,	
	caption={Example Database Schema},
	label={dl:set_union_inject},
	commentstyle=\color{red},
	%float=tp,
	%floatplacement=tbp,
	%belowskip=-.5em
]
  <\textbf{ontime}>(<\textcolor{purple}{fid}>,<y>,<m>,<d>,<h>,adelay,ddelay,<\textcolor{orange}{src\_apid}>,<\textcolor{orange}{dst\_apid}>,<\textcolor{darkgreen}{alid}>)
  <\textbf{airlines}>(<\textcolor{darkgreen}{alid}>,name,active)
  <\textbf{airports}>(<\textcolor{orange}{apid}>,name,lat,lon,elevation,city,<\textcolor{blue}{state}>,country)
  <\textbf{shapes}>(<\textcolor{blue}{state}>, polygons[])
\end{lstlisting}
}

\noindent The \texttt{ontime} table records the arrival and departure delays of each flight (i.e., \texttt{adelay} and \texttt{ddelay}, respectively) from a source airport with id \texttt{\textcolor{orange}{src\_apid}} to a destination airport with id \texttt{\textcolor{orange}{dst\_apid}} along with the departure time of the flight (i.e., \texttt{\textbf{y}ear}, \texttt{\textbf{m}onth}, \texttt{\textbf{d}ay}, and \texttt{\textbf{h}our}) and the carrier that operated the flight \texttt{\textcolor{darkgreen}{alid}}. The \texttt{airports} table records the id of each airport (\texttt{\textcolor{orange}{apid}}) along with its \texttt{name},  latitude (\texttt{lat}), longitude (\texttt{lon}), \texttt{elevation}, \texttt{city}, \texttt{\textcolor{blue}{state}}, and \texttt{country}. The \texttt{airlines} table stores the id of an airline (\texttt{\textcolor{darkgreen}{alid}}) along with its \texttt{name} and whether or not the airline is still \texttt{active}.  \texttt{airports} and \texttt{airlines} serve as dimensions tables to the fact table \texttt{ontime}. Finally, the \texttt{shapes} table records an array of \texttt{polygons} that corresponds to the geographical bounds of each \texttt{\textcolor{blue}{state}} in the US.

\begin{figure*}[tbh]
  \centering
  \begin{subfigure}[t]{.4\textwidth}
    \centering
    \includegraphics[width=\columnwidth]{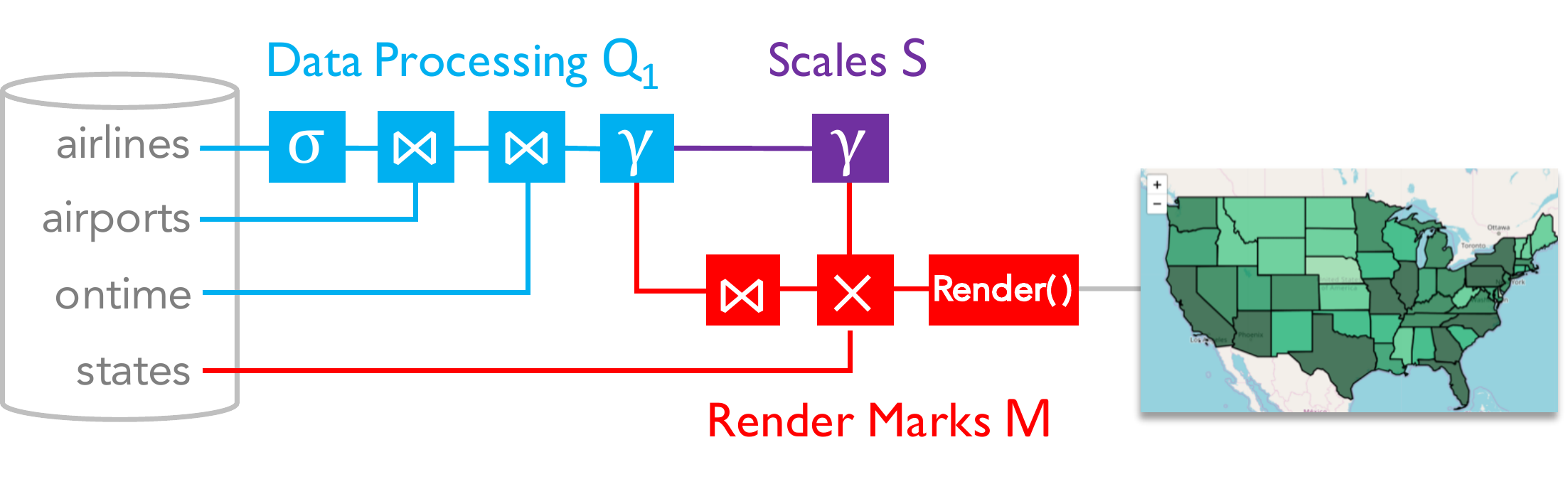}
    \caption{Static visualization.}
    \label{f:vis1_real}
  \end{subfigure}
	\hfill
  \begin{subfigure}[t]{.28\textwidth}
    \centering
    \includegraphics[width=\textwidth]{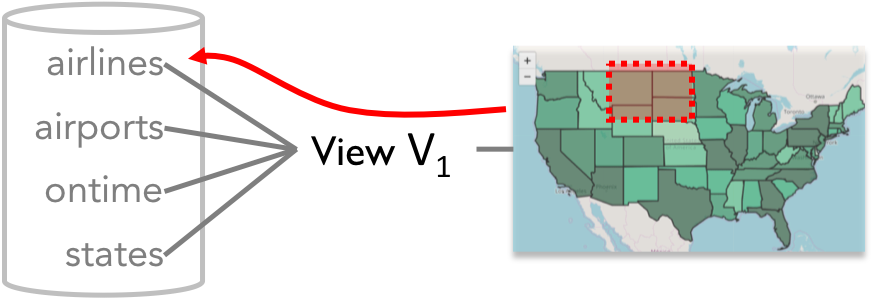}
    \caption{Selection interaction.}
    \label{f:vis2}
  \end{subfigure}
	\hfill
  \begin{subfigure}[t]{.28\textwidth}
    \centering
    \includegraphics[width=\textwidth,clip,trim=.4em 0 .4em .2em]{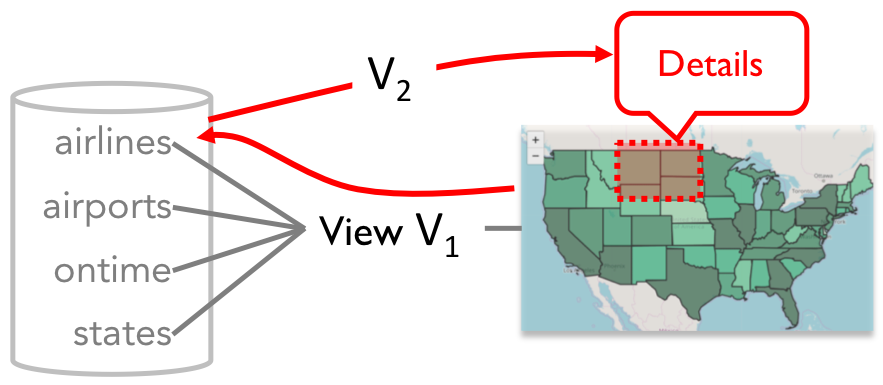}
    \caption{Tooltip/Details-on-demand.}
    \label{f:vis3}
  \end{subfigure}
  \caption{(a) breaks down a visualization view $V_1$ into \textcolor{blue}{data processing}, \textcolor{magenta}{value range computation}, and \textcolor{red}{mark rendering} operators. (b) shows the logical backward trace operation over $V_1$ to identify the subset of \texttt{ontime} tuples that contribute to an interactive range selection. (c) shows how using the identified subset in another view $V_2$ can be used to show details for this selection.}
  \label{}
\end{figure*}

\stitle{Initial Static Visualization.} Let us start by building a static visualization to depict the number of flights for active airlines per state as a heatmap, similar to the one in~\Cref{fig:intro}(A). In purely relational terms, we can specify this visualization as follows:
{\small
\begin{lstlisting}[
	language = SQL,
	showspaces=false,
	basicstyle=\ttfamily\footnotesize,
	commentstyle=\color{gray},
	mathescape=true,
	numbers=none,
	frame = single,
	escapeinside={<}{>},
	captionpos=b,	
	caption={Example of a static visualization.},
	label={dl:q1},
	commentstyle=\color{red},
	%float=tp,
	%floatplacement=tbp,
%	belowskip=-.5em
]
-- Data Processing
Q1 = SELECT COUNT(*) AS cnt, 
            AVG(ddelay) AS avg_ddelay,
            AVG(adelay) AS avg_adelay, 
            state
     FROM   ontime, airports, airlines
     WHERE  ontime.alid = airlines.alid AND
            ontime.src_apid = airports.apid AND
            airlines.active = 'Y'
     GROUP BY state
-- Visualization
S = SELECT MIN(cnt) AS mi, MAX(cnt) AS mx FROM Q1
M = SELECT states.polygons,         -- geometry
           color(Q1.cnt,S.mi,S.mx)  -- color
    FROM   Q1, S, states
    WHERE  states.state = Q1.state
P = render_map(M)
\end{lstlisting}
}
\noindent\Cref{f:vis1_real} depicts the workflow described by the above queries.  \texttt{Q1} specifies the data processing part of the visualization and consists of a join between the \texttt{ontime}, \texttt{airlines}, and \texttt{airports} relations followed by a filter on active airlines and a group-by state count aggregation. (\texttt{Q1} also computes the average departure and arrival delays per state that we use later in interactions.) \texttt{M} constitutes part of the visualization workflow that transforms the output of \texttt{Q1} into attributes of polygon marks (i.e., geometry and color of each polygon).   \texttt{color()} is syntactic sugar for an equation that maps each count value to an output range of green hues, where the input range is computed by \texttt{S} as the minimum and maximum counts from \texttt{Q1}.   Finally, the polygons are rendered on the screen using a mark-specific \texttt{render\_map()} shim.  We omit further details for space considerations and refer interested readers to prior work in relational specifications of visualization workflows~\cite{wu2017dvms,EugeneWuVision2014}.  

Under this model, the visualization application is a (possibly complex) relational view $V_i$ that maps the input database in \emph{data space} to rendered marks in \emph{pixel space}.  Next, we elaborate on the connections of common interactive capabilities and provenance concepts by building on this static visualization example.

\sstitle{Interactive Selections.} One of the fundamental building blocks of visualization management systems is the ability to interactively reference visual marks by clicking, lassoing, or other types of selection operations~\cite{tukey1977exploratory,satyanarayan2017vegalite,wilhelm2003taxonomy}.      Although users interact with visual marks, the intention is typically to manipulate the underlying data represented by the visual marks rather than the marks themselves.\footnote{Note that this is not strictly always the case.  For instance, users may want to reconfigure marks (e.g., change their color)  without referencing base data~\cite{yi2007taxonomy,harper:2014:d3constructing}.} To this end, visualization research has developed many techniques to invert selections in pixel space to declarative selection queries in the input data space~\cite{satyanarayan2017vegalite,heer2008generalized,vqe1997derthick,livny1997devise, north2000snap}. 

The predominant forms of selection are item/group selection and range selection.  Consider the map in~\Cref{f:vis2}. Item and group selection may correspond to clicking on one or more states, where the selection is a set of states.  The intention is to identify the input records associated with the selected states.  Range selection may correspond to drawing a bounding box (dashed \red{red} box).  This may be interpreted as group selection, where the set of states corresponds to the state polygons that intersect with the box.  However, the intention may also be to translate the bounding box into a predicate over \texttt{lat,lon} attributes in the shapes polygons.  The latter representation can be attractive because the selection can be further manipulated and relaxed to, say, add additional predicates (e.g., \texttt{adelay} > $5min$), modify the predicate clauses (e.g., increase the \texttt{lon} range), or remove unnecessary clauses~\cite{heer2008generalized}.

\ititle{Connection with Provenance:}  All of the above selection types are variants of a common provenance operation known as \emph{backward trace}, which identifies input records that contribute to specified output records.  Different backward trace implementation techniques correspond to the above selection semantics.   

Visualization systems typically support range selection when the visualization workflow consists of rescaling data attributes to visual variables (e.g., \texttt{COUNT} to y pixel position).  Since the scaling operations are typically invertible, it is simple to, say , rescale a bounding box's coordinates from $y_{min}$ to $y_{max}$ to be in terms of \texttt{COUNT}.   Provenance research generalizes this by computing the workflow's inverse function $V_i^{-1}()$.  This can be done through weak inverse functions~\cite{woodruff1997supporting}, deriving provenance predicates from relational workflows~\cite{ikedathesis}, or by explicitly annotating each operator with an inverse function~\cite{wu2013subzero}.  Expressing range selections as backward trace helps extend its support to visualizations that perform complex data processing, as well as rendering. 

Item (or group) selections identify the specific input records that correspond to the user's selection in pixel space.  Visualization systems typically implement this by annotating records as they flow through the visualization workflow so that the output is annotated with the input records~\cite{Bostock11d3}.  However, annotations~\cite{dbnotes2004bhagwat,gprom} are only one mechanism to answer fine-grained provenance queries.  They can also be computed  by evaluating the provenance predicates above, or by explicitly materializing input-to-output record dependency information as explicit index data structures when executing the visualization workflow~\cite{wu2013subzero,psallidas2018smoke}.

\ititle{A note on semantics:} One subtle point is that provenance systems may support different types of provenance semantics, and visualization developers should be aware of these semantics.  For instance, assume we select outputs of Q1 and want the corresponding airlines from the \texttt{airlines} relation. We typically only want the set of airlines, rather than the bag of every copy of the airlines that were used to derive the selection. In this case, visualization toolkits should demand ``which-provenance'' semantics~\cite{tannen2017pods} as opposed to general transformation provenance semantics that return each airline tuple as many times as it contributes to selected outputs. (See~\cite{tannen2017pods,provdb,ikedathesis} for an introduction to different provenance semantics.)

\stitle{Tooltips and Details-on-Demand.}  A common use case once a user has performed a selection is to show detailed information, or summarizations, about the selected data. Tooltips and details-on-demand are popular examples of this paradigm.  

Tooltips render information (say, in a modal pop-up) that contains information about the provenance of the selected marks. For instance, when users select states in~\Cref{f:vis1_real}, they may want to see additional attributes per state such as the average arrival and departure delays (i.e., \texttt{avg\_adelay} and \texttt{avg\_ddelay}, respectively).  

Details-on-demand go beyond tooltips by retrieving and further processing user selections. For instance, when hovering over a state, the visualization may update to show a detailed list of airports operating in the state. Another form of details-on-demand is to semantically zoom into the user's range selection.  For instance, the user may select states with a range selection on the map. In response, the visualization updates to zoom into the range and show, say, detailed \emph{city-}level breakdowns of counts of delayed flights.

\ititle{Connection with Provenance:} These functionalities are often implemented as standalone features in a visualization system. However, they can be easily expressed as queries that take the backward trace of the user's selection as input.  We illustrate this in \Cref{f:vis3}.  The user selection in the visualization is traced back to the input records, then a second visualization workflow $V_2$ (often expressed as a SQL query) computes statistics about the provenance and renders them as details. The primary distinction between the above examples is the definition of $V_2$, which we illustrate in~\Cref{dl:tdod} below:    
{\small
\begin{lstlisting}[
	language = SQL,
	showspaces=false,
	basicstyle=\ttfamily\footnotesize,
	commentstyle=\color{red},
	mathescape=true,
	numbers=none,
	frame = single,
	escapeinside={<}{>},
	captionpos=b,
	caption={Examples of tooltips and details-on-demand},
	label={dl:tdod},
%	commentstyle=\color{red},
	%float=tp,
	%floatplacement=tbp,
%	belowskip=-.5em
]
-- Tooltip
T = SELECT   avg_ddelay, avg_adelay
    FROM     backward_trace(selected, Q1)
    
-- Details-on-demand
D = SELECT * FROM backward_trace(selected, airports);
Z = SELECT   COUNT(*), city
    FROM     backward_trace(selected, ontime) A1,
             backward_trace(selected, airports) A2
    WHER     A1.alid = A2.alid
    GROUP BY city;
\end{lstlisting}
}
\noindent The tooltip query \texttt{T} traces the provenance of the user's \texttt{selected} states to the output of \texttt{Q1}, and returns the average departure and arrival delays.  The details-on-demand shows two queries.  \texttt{D} retrieves the list of airports within the selected states.   \texttt{Z} performs the drill-down from state to city-level statistics, for the selected states.  It does this by joining \texttt{ontime} records and airports for the selected states, and re-computes the number of delays for each city.

\ititle{A note on performance: } Joins, such as the one in the query \texttt{Z} above, are common in visualizations.  To avoid potentially expensive join execution costs,  it is common practice for visualization systems and developers to first denormalize relations ahead of time.  The visualization is then implemented over the denormalized relation.  

However, denormalization is only one possible join optimization and comes with several costs.  It introduces redundancy, is time- and space-consuming to construct, and in many cases not even required.  Furthermore, this focus on denormalization is an example of violating physical data independence~\cite{codd} and impedes rapid visualization development.   For instance, developers may spend considerable time writing application code to essentially denormalize \texttt{ontime}$\bowtie$\texttt{airports} and compute the per-city count.   Later, they may want to iterate on the visualization design and try showing, say, other statistics or grouping by \texttt{elevation}. However, they may be reluctant to incur the same engineering cost to try another design. This is because each design change implies the time- and space-consuming process of reconstructing the denormalized relation.  

In contrast, expressing this logic in terms of provenance and relational operations enables rapid design iteration  by offloading implementation to the visualization engine.  Furthermore, recent work~\cite{psallidas2018smoke} suggests that workflows composed of provenance and relational operations can be optimized to ensure interactive response times by, say, materializing efficient join indexes adaptively, partially denormalizing the database, and pre-computing statistics.

\begin{figure}[h]
  \begin{subfigure}[t]{.475\columnwidth}
    \centering
    \includegraphics[width=\textwidth]{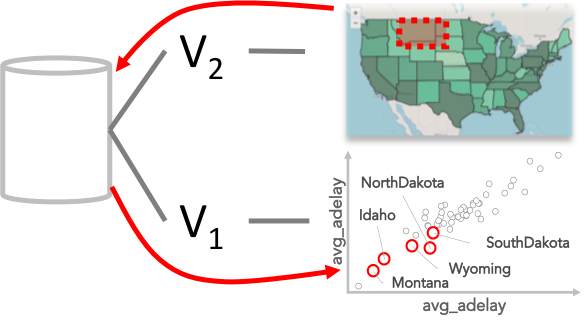}
    \caption{}
    \label{f:vis4_link}
  \end{subfigure}
	\hfill
  \begin{subfigure}[t]{.475\columnwidth}
    \centering
    \includegraphics[width=\textwidth,trim={.4em 1.5em .4em .4em},clip]{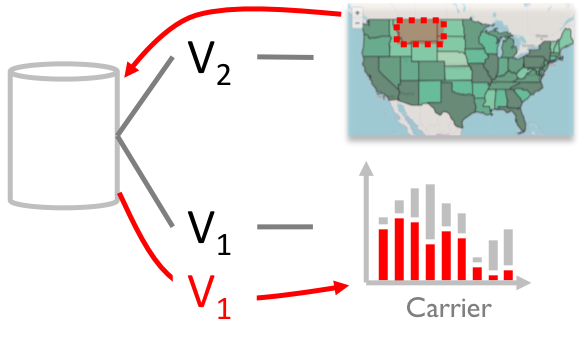}
    \caption{}
    \label{f:vis4_cross}
  \end{subfigure}
  \caption{Linking and Cross-filtering}
  \label{f:vis4}
\end{figure}

\stitle{Multi-View Linking.} Linking is a common class of interactions where selections in one view update other views. Prominent examples include linked brushing and cross-filtering.

\ititle{Linked brushing:}  Suppose we render a scatterplot of the average arrival (y-axis) and departure (x-axis) delays for each state, as computed by \texttt{Q1} in~\Cref{dl:q1}. Consider the visualization in~\Cref{f:vis4_link}. Linked brushing may let users select states on the map (\red{red} box) to highlight the corresponding delay information for each selected state in the scatter plot (\red{red} circles), and vice versa. 

\ititle{Cross-filtering:} Cross-filtering is used to explore correlated statistics across multiple visualization views~\cite{crossfilter}.   In the common setup, each view is the result of an aggregation query over different combinations of input attributes (e.g., each view in Figure~\ref{fig:intro}).  Selecting marks in one view recomputes the aggregation queries over the subset of input records represented by the selection, and updates the views accordingly.   \Cref{f:vis4_cross} illustrates a simple example where selecting a set of states updates the counts of flights per carrier.

\ititle{Connection to Provenance:}  Linked brushing is precisely backward tracing from the states to the input state records, followed by forward tracing to highlight the states in the scatterplot.   Cross-filtering is expressed as backward tracing followed by refreshing the other views by executing the queries (e.g., $V_1$ in Figure~\ref{f:vis4_cross}) over the provenance.    The difference is based on the forward tracing operation. In this example, linked brushing traces the subset to the output marks, whereas cross-filtering recomputes the views for the output marks.

\ititle{A note on semantics:} To better highlight the importance of the provenance literature in the domain of interactive applications, note that the update procedure corresponds to a common provenance operation, known as \emph{selective refresh} in the provenance literature. Selective refresh may not always update the same target outputs, for instance if the workflow contains a one-to-many operator followed by two non-monotonic aggregation operators~\cite{ikedathesis}.  The notion of unsafe selective refresh, and recent techniques to address it~\cite{chothia:2016:explaining}, highlight the value of leveraging  the provenance literature to ensure correctness in interactive visualizations. 

\ititle{A note on performance: } Crossfilter is an important yet computationally expensive interaction technique. The visualization community has begun adopting dense~\cite{liu2013immens} and sparse~\cite{lins2013nanocubes} data cubes to support cross-filtering at interactive speeds. Unfortunately, building such data structures requires considerable offline time--from minutes to hours on the ontime flights dataset.  This ``cold-start'' problem~\cite{position:2017:leilani} makes it challenging for developers to rapidly build and test complex interactive visualizations, and makes it difficult to load a dataset in a visualization engine and immediately start cross-filtering.  

Recent work~\cite{psallidas2018smoke} on fast fine-grained provenance engines shows that it is possible to construct whole or partial data cubes for cross-filter provenance queries in interactive time.    In addition, provenance metadata can be represented in efficient index data structures that accelerate backward and forward provenance tracing lookups.  These forward and backward indexes are precisely the indexes to support incremental view updates on deletion.

\section{Provenance-Supported Interaction}
\label{s:crazy}

Section~\ref{s:express} described how core visualization interactions can be succinctly expressed in terms of provenance.  This means that a visualization engine that is engineered to support provenance querying can readily add support for such interactions. Developers can then declaratively specify interactive visualizations and rely for their optimization on the underlying provenance-enabled visualization engine. In this section, we look beyond existing interactive visualization features, and examine {\it new functionality that may be possible} with the capabilities of such a provenance-enabled engine.

\begin{figure}[t]
  \includegraphics[width=\columnwidth]{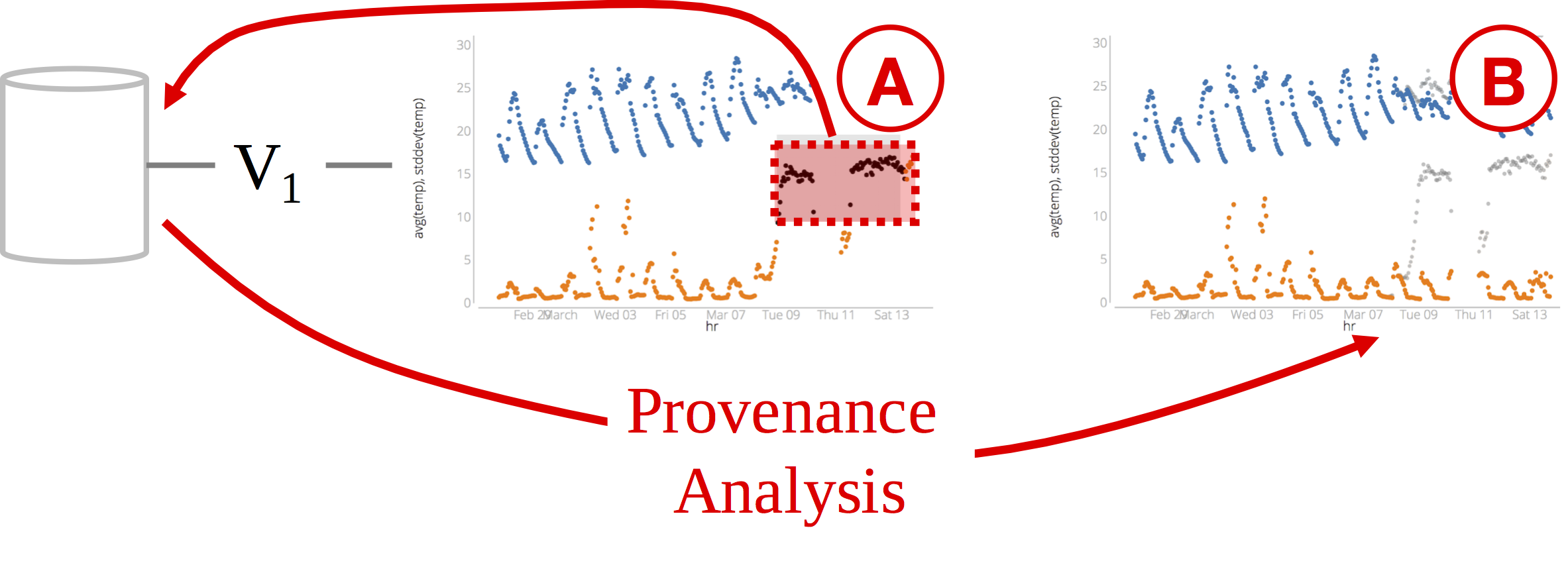}
  \caption{Before and after of an advanced provenance analysis.   (a) the user selects outliers in the initial visualization (shown on the left), and (b) the results of the predicate explanation update the visualization (shown on the right).  In practice, the visualization will update in place.}
  \label{f:vis7}
\end{figure}

\stitle{Advanced Provenance Analysis.} To begin, we first highlight a rich area of provenance analysis techniques, such as interactive query specification~\cite{dataplay2012abouzied}; what-if analysis~\cite{provisioning2015assadi,deutch2013caravan}; and result explanation~\cite{scorpion} among others, that already exists.  These techniques are a natural fit with a provenance-enabled visualization engine (\Cref{f:vis7}).  First, their inputs consist of provenance metadata and user-provided information that can be naturally elicited through a visualization interface.  Second, their outputs are often in the form of predicates, records, or queries that can be naturally rendered in a visualization. Furthermore, they can be integrated as a function over the provenance result in a similar way to cross-filtering in \Cref{f:vis4_cross}. We illustrate a few examples of such integration below.

\ititle{Data Explanation: } Outlier explanation techniques~\cite{scorpion,roy2015explain,wu2012demonstration} take as input anomalies in the visualized data, the query used to generate the visualization, and return simple predicates that are most ``responsible'' for those errors.   \Cref{f:vis7} shows how this is integrated into an interactive visualization.  The user selects anomalies in the scatter plot on the left (A). Then, the analysis procedure uses $V_1$ and the fine-grained provenance of the selected points to generate a predicate explanation.  Rather than print the explanation in textual form, it can be deeply integrated into the visualization itself.  The example visualization recomputes the query $V_1$ over a subset of the input identified by the explanation and renders it as an overlay (B).

\ititle{Why-not Analysis: } Non-existence of anticipated query results play a detrimental role on the overall data exploration and analysis. For instance, if the state of California was missing in the map plot of~\Cref{fig:intro}, then the user may be confused.  Similarly, if the user complains that the \texttt{COUNT} of delayed flights should be higher for a specific carrier (perhaps by resizing a bar to be higher), then the user is questioning the absence of delayed flights in the visualization. Although the algorithms for generating these explanations~\cite{gpromwhynot2017lee,dataplay2012abouzied} may differ, the way they can be integrated into, and presented within, the visualization are similar to the preceding example.

\stitle{Multi-application Linking.}
Visualizations contain multiple views in order to present patterns between important combinations of attributes (\Cref{fig:intro}).  Cross-view interactions such as linked brushing and cross-filtering are powerful because they help the user identify relationships between patterns.

 In terms of functionality, they combine record-level backward tracing from selections in the visualization  with forward tracing to (and refreshing of) visualizations dependent on shared input data.  By expressing these interactions in terms of provenance it becomes clear that the backward and forward tracing operations need not be coupled, {\it nor even be implemented within the same visualization application}.  As long as different applications process the same dataset, and support backward and/or forward tracing functionality, then linking and cross-filtering {\it across multiple applications} is possible.

\begin{figure}[t]
  \includegraphics[width=.95\columnwidth]{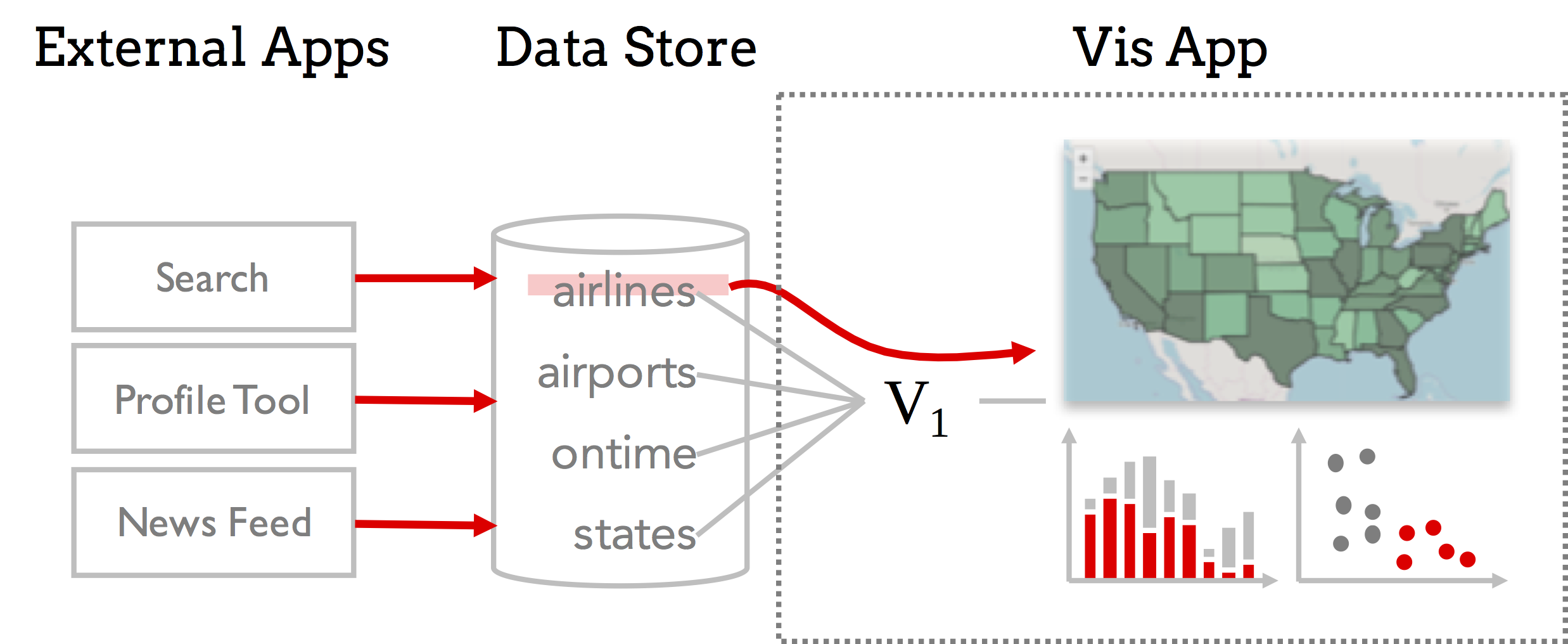}
  \caption{Provenance can enable linking and cross-filtering across different applications.}
  \label{f:vis6}
\end{figure}

\begin{figure*}[tb]
  \begin{subfigure}[t]{.31\textwidth}
    \centering
    \includegraphics[width=\textwidth]{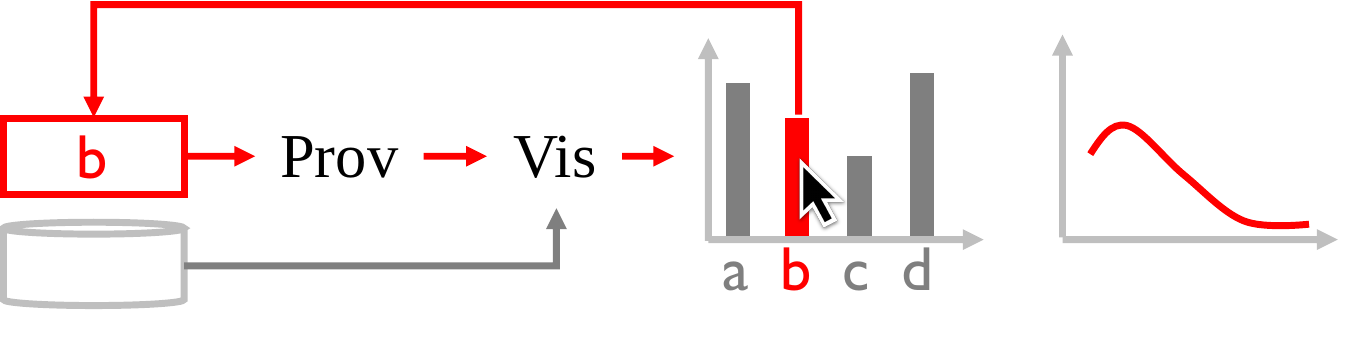}
    \caption{Hovering over bar \red{\texttt{b}} triggers an interaction event to trace \red{\texttt{b}}'s provenance  ($Prov$) and update the line chart  ($Vis$).}
    \label{f:vis5_b}
  \end{subfigure}\hfill
  \begin{subfigure}[t]{.31\textwidth}
    \centering
    \includegraphics[width=\textwidth]{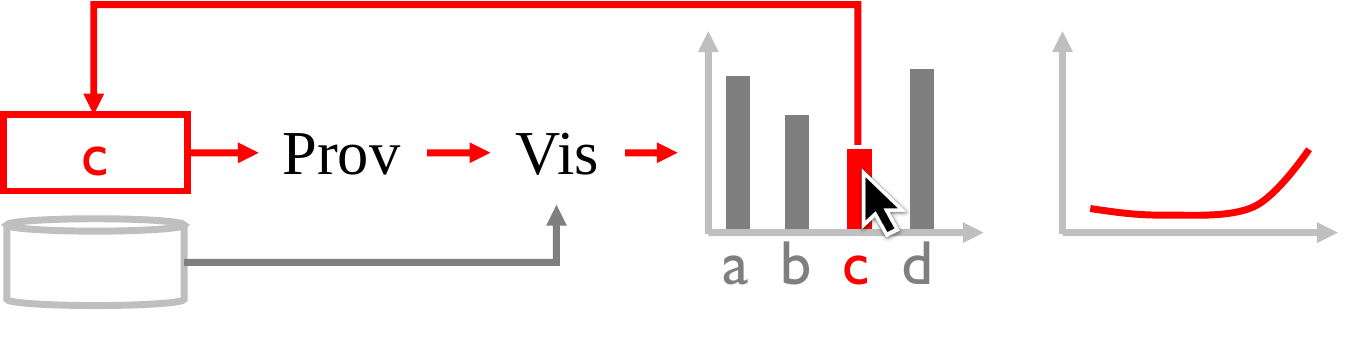}
    \caption{Hovering over bar \red{\texttt{c}} performs the same logic but for the event associated with \red{\texttt{c}}. }
    \label{f:vis5_c}
  \end{subfigure}\hfill
  \begin{subfigure}[t]{.31\textwidth}
    \centering
    \includegraphics[width=\textwidth]{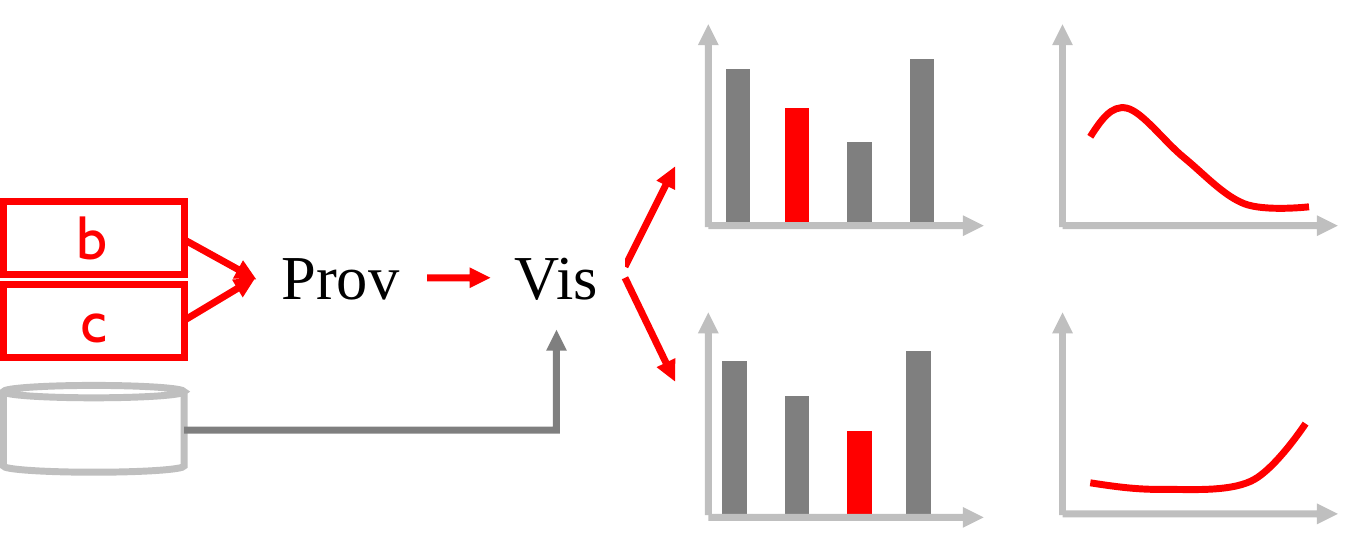}
    \caption{Explicitly tracking the provenance as a relation of events can easily render a history of past events.   }
    \label{f:vis5_both}
  \end{subfigure}
  \caption{Provenance of a cross-filter interaction can be modeled as the history of the visualization's interaction events.} \label{f:vis5}

\end{figure*}

Figure~\ref{f:vis6} illustrates linking between the running visualization example with external applications such as search and user profile management.  The user may use a form-based search interface to find recent flights through Miami. This result set is fundamentally the result of a query workflow over the data store but presented as a text- and image-based web application.  By tracing these search results back to the input data (the \red{red} rectangle over \texttt{airlines} represents a subset of the relation), they can also be traced forward to update the visualization application (depicted by the \red{red} arrows). Furthermore, changing the search parameters updates both the search results and the visualization.  The reverse is also possible: selecting data in the visualization can also update the search results.   

Similarly, user profile tools that show to users their past flights and bookings can be linked to update the visualization to show delay statistics of the user's past flights, as well as to update the search results with flights the user has taken.  In short, any application that tracks backward provenance can issue interactions that update the presentation in any application that supports forward provenance, as long as the two ends coordinate on the same base relations.

\stitle{Provenance of Interactions.}
So far, we have described how provenance can be {\it used} to express the results of interactions.   For example, Figure~\ref{f:vis4}(b) shows that the bottom bar chart is updated by re-running $V_1$ over the backward provenance of the highlighted bars in the top bar chart.  In many cases, interactions simply change the inputs to the application logic (e.g., $V_1$, $V_2$) rather than the logic itself.  In these cases, interactions are a form of input data, whose provenance and versions can be tracked.  

Figure~\ref{f:vis5} illustrates this for a simple cross-filtering visualization, where hovering over bars in the bar chart updates the line chart.  We have simplified the workflow for clarity.  $Vis$ describes all application logic to compute and render both views; it is analogous to the union of $V_1$ and $V_2$ in Figure~\ref{f:vis4}.  When the user hovers over the \red{\texttt{b}} bar, the cross-filter logic executes $Vis(Prov(\redtt{b}))$ to update the visualization (shown as the \red{red} arrows in \Cref{f:vis5_b}).  The cross-filter logic is typically written within an event-handler that executes for each interaction event.\footnote{In a relational context, where the visualization is modeled as a materialized view, this is similar to scheduling view updates in response to changes in input relations.} Thus, when the user hovers over bar \redtt{c}, the cross-filter logic simply executes $Vis(Prov(\redtt{c}))$, shown in \Cref{f:vis5_c}.

Note that the interaction events \redtt{b} and \redtt{c} are data, thus we might track the provenance of the visualization interactions in e.g., a relation of \texttt{events} (\Cref{f:vis5_both} shows a relation containing \redtt{b,c}).   This relation lets us decouple visualization update logic from user interactions, and manage them explicitly.  For instance, \Cref{f:vis5_both} shows how a history of past events can be presented and ~\Cref{dl:q4} shows how it can be implemented. Similarly, selecting a single record is akin to undo or time-travel. Advanced functionality may select a 2D-range of marks, and query for historical {\it interactions} (backward provenance to the \texttt{events} relation) that generated charts based on the selection (forward provenance to historical visualizations).
 {\small
 \begin{lstlisting}[
 	language = SQL,
 	showspaces=false,
 	basicstyle=\ttfamily\footnotesize,
 	commentstyle=\color{gray},
 	mathescape=true,
 	numbers=none,
 	frame = single,
 	escapeinside={<}{>},
 	captionpos=b,	
 	caption={Query pseudocode to render history of interactions generated from the bar chart.},
 	label={dl:q4},
 	commentstyle=\color{red},
 	%float=tp,
 	%floatplacement=tbp,
 %	belowskip=-.5em
 ]
     SELECT Vis(Prov(e)) FROM events e 
     WHERE  e.source = 'barchart';
 \end{lstlisting}
 }

\stitle{Application Design Search.} Tracking~\cite{hellerstein2017ground} and recovering~\cite{mavlyutov2017dependency,halevy2016goods} coarse-grained provenance in order to understand how workflows and applications throughout an organization result in reads and writes of data files.  This can be helpful if a developer wants to analyze a given dataset, by suggesting previous workflows that have processed the same files.  Similar functionality can help provide inspiration for visualization and application developers.   For example, visualization developers that want to analyze flight delays for the North American marketing team can use coarse-grained provenance to find visualizations that use the flight relations.  They can use these visualizations, such as \Cref{fig:intro}, to interactively specify the subset of the flight relations they want to work with.  Based on this subset of records, fine-grained provenance can be used to identify the visualizations that primarily uses this specific subset.  This iterative form of refinement can help the developer find the most relevant designs and application logic to borrow from, or perhaps find that their desired visualization already exists.

\stitle{Interaction-By-Example.} View synthesis and query-by-example systems~\cite{psallidas2015s4,mottin2014eqg} address the problem where, given an input database and examples of desired query results, the goal is to return queries that generate the example results (or a superset).  This formulation can be attractive because SQL queries are known to be hard to compose.  However, the general problem is very challenging due to the expressiveness of SQL, and approaches typically focus on a semantically meaningful subset of the language for which identifying the queries by output examples can be efficient.

Earlier, we described how a wide range of visualization interactions can be decomposed into combinations of provenance queries.  Thus, there is potential to develop {\it interaction-by-example}, where the user  directly selects and manipulates parts of a static visualization  (e.g., drag marks to new locations) to specify an example of a desired interaction.  This is akin to~\cite{scheidegger2007querying} but specific to fine-grained data visualization lineage rather than coarse-grained workflow provenance. A synthesis engine can then generate the appropriate provenance statements to support the interaction.  The simplicity of provenance queries---namely coarse-grained and fine-grained backward and forward queries, along with refresh---suggests that this may be both tractable and semantically meaningful.

\stitle{Deconstruction and Restyling.}  Harper et al.~\cite{harper:2014:d3constructing} present a technique to extract data from marks in D3 visualizations and re-style the data using new visual encodings.  For instance, a bar chart might be restyled into a scatterplot that is colored differently.  Their technique relied on D3 because it automatically annotates each mark with the record used to generate the mark.  However, D3 does not track annotations across data processing workflows, thus restyling is limited to design.   In contrast, tracking provenance can let users restyle the data processing, for example by plotting \texttt{MAX} rather that \texttt{COUNT} statistics, or modifying the semantics of linked interactions. \eat{Provenance-based techniques that present to the user such meaningful re-styling automatically are interesting future work.}   
\section{Discussion}
\label{s:conclusions}

Provenance is a fundamental type of information with wide applications across domains.  In this paper, we showed that provenance can serve as the logical underpinning of well-established, as well as novel, interactive visualization functionalities.  Overall, the purpose, and corresponding takeaways, of this paper is three-fold:

First, is to convey the value of leveraging provenance capabilities and semantics to declaratively express and design visualization applications.   Current visualization developers build custom data structures and make optimization choices that are coupled with interactions that can be efficiently supported; changing the visualization interactions often means rearchitecting the entire visualization application.   Expressing interactive visualizations in terms of provenance introduces physical data independence, and can help developers rapidly iterate upon visualization designs. 

Second, is to highlight the need for fast coarse- and fine-grained provenance engines.  Traditionally, data processing engines that support fine-grained provenance expect to incur non-trivial amounts of overhead in order to quickly answer provenance queries, yet interactive visualizations are only useful if the application responds within interactive latencies.   Recent work~\cite{psallidas2018smoke,psallidas2018smokedemo} showed evidence that fine-grained provenance can both be materialized at interactive speeds, and be used as index data structures to {\it accelerate} visualization queries.    We believe that the connections between query optimization and provenance is a rich area of research that is worthy of further pursuit. 

Finally, interactive visualizations are a prominent type of data-driven interactive applications.  We believe many of the connections and benefits described in this paper can translate to the general class of optimizing and expressing interactive applications.

\balance

\bibliographystyle{ACM-Reference-Format}
\bibliography{main}


\begin{thebibliography}{64}


\ifx \showCODEN    \undefined \def \showCODEN     #1{\unskip}     \fi
\ifx \showDOI      \undefined \def \showDOI       #1{#1}\fi
\ifx \showISBNx    \undefined \def \showISBNx     #1{\unskip}     \fi
\ifx \showISBNxiii \undefined \def \showISBNxiii  #1{\unskip}     \fi
\ifx \showISSN     \undefined \def \showISSN      #1{\unskip}     \fi
\ifx \showLCCN     \undefined \def \showLCCN      #1{\unskip}     \fi
\ifx \shownote     \undefined \def \shownote      #1{#1}          \fi
\ifx \showarticletitle \undefined \def \showarticletitle #1{#1}   \fi
\ifx \showURL      \undefined \def \showURL       {\relax}        \fi
\providecommand\bibfield[2]{#2}
\providecommand\bibinfo[2]{#2}
\providecommand\natexlab[1]{#1}
\providecommand\showeprint[2][]{arXiv:#2}

\bibitem[\protect\citeauthoryear{Abouzied, Hellerstein, and
  Silberschatz}{Abouzied et~al\mbox{.}}{2012}]%
        {dataplay2012abouzied}
\bibfield{author}{\bibinfo{person}{Azza Abouzied}, \bibinfo{person}{Joseph
  Hellerstein}, {and} \bibinfo{person}{Avi Silberschatz}.}
  \bibinfo{year}{2012}\natexlab{}.
\newblock \showarticletitle{DataPlay: Interactive Tweaking and Example-driven
  Correction of Graphical Database Queries}. In
  \bibinfo{booktitle}{\emph{Proceedings of the 25th Annual ACM Symposium on
  User Interface Software and Technology}} \emph{(\bibinfo{series}{UIST '12})}.
\newblock


\bibitem[\protect\citeauthoryear{Alabi and Wu}{Alabi and Wu}{2016}]%
        {alabipfunk}
\bibfield{author}{\bibinfo{person}{Daniel Alabi} {and} \bibinfo{person}{Eugene
  Wu}.} \bibinfo{year}{2016}\natexlab{}.
\newblock \showarticletitle{{PFunk-H}: Approximate Query Processing using
  Perceptual Models}. In \bibinfo{booktitle}{\emph{Proceedings of the Workshop
  on Human-In-the-Loop Data Analytics}} \emph{(\bibinfo{series}{HILDA '16})}.
\newblock


\bibitem[\protect\citeauthoryear{Althoff, Dong, Murphy, Alai, Dang, and
  Zhang}{Althoff et~al\mbox{.}}{2015}]%
        {althoff2015timeline}
\bibfield{author}{\bibinfo{person}{Tim Althoff}, \bibinfo{person}{Xin~Luna
  Dong}, \bibinfo{person}{Kevin Murphy}, \bibinfo{person}{Safa Alai},
  \bibinfo{person}{Van Dang}, {and} \bibinfo{person}{Wei Zhang}.}
  \bibinfo{year}{2015}\natexlab{}.
\newblock \showarticletitle{TimeMachine: Timeline Generation for Knowledge-Base
  Entities}. In \bibinfo{booktitle}{\emph{Proceedings of the 21th ACM SIGKDD
  International Conference on Knowledge Discovery and Data Mining}}
  \emph{(\bibinfo{series}{KDD '15})}.
\newblock


\bibitem[\protect\citeauthoryear{Assadi, Khanna, Li, and Tannen}{Assadi
  et~al\mbox{.}}{2015}]%
        {provisioning2015assadi}
\bibfield{author}{\bibinfo{person}{Sepehr Assadi}, \bibinfo{person}{Sanjeev
  Khanna}, \bibinfo{person}{Yang Li}, {and} \bibinfo{person}{Val Tannen}.}
  \bibinfo{year}{2015}\natexlab{}.
\newblock \showarticletitle{Algorithms for provisioning queries and analytics}.
\newblock \bibinfo{journal}{\emph{CoRR}}  \bibinfo{volume}{abs/1512.06143}
  (\bibinfo{year}{2015}).
\newblock


\bibitem[\protect\citeauthoryear{Battle, Chang, Heer, and Stonebraker}{Battle
  et~al\mbox{.}}{2017}]%
        {position:2017:leilani}
\bibfield{author}{\bibinfo{person}{Leilani Battle}, \bibinfo{person}{Remco
  Chang}, \bibinfo{person}{Jeffrey Heer}, {and} \bibinfo{person}{Michael
  Stonebraker}.} \bibinfo{year}{2017}\natexlab{}.
\newblock \showarticletitle{Position Statement: The Case for a Visualization
  Performance Benchmark}. In \bibinfo{booktitle}{\emph{Proceedings of the 2nd
  Workshop on Data Systems for Interactive Analysis}}
  \emph{(\bibinfo{series}{DSIA '17})}.
\newblock


\bibitem[\protect\citeauthoryear{Battle, Chang, and Stonebraker}{Battle
  et~al\mbox{.}}{2016}]%
        {battle2016}
\bibfield{author}{\bibinfo{person}{Leilani Battle}, \bibinfo{person}{Remco
  Chang}, {and} \bibinfo{person}{Michael Stonebraker}.}
  \bibinfo{year}{2016}\natexlab{}.
\newblock \showarticletitle{Dynamic prefetching of data tiles for interactive
  visualization}. In \bibinfo{booktitle}{\emph{Proceedings of the 2016
  International Conference on Management of Data}}
  \emph{(\bibinfo{series}{SIGMOD '16})}.
\newblock


\bibitem[\protect\citeauthoryear{Bhagwat, Chiticariu, Tan, and
  Vijayvargiya}{Bhagwat et~al\mbox{.}}{2004}]%
        {dbnotes2004bhagwat}
\bibfield{author}{\bibinfo{person}{Deepavali Bhagwat}, \bibinfo{person}{Laura
  Chiticariu}, \bibinfo{person}{Wang~Chiew Tan}, {and} \bibinfo{person}{Gaurav
  Vijayvargiya}.} \bibinfo{year}{2004}\natexlab{}.
\newblock \showarticletitle{An Annotation Management System for Relational
  Databases}. In \bibinfo{booktitle}{\emph{Proceedings of the 30th
  International Conference on Very Large Data Bases}}
  \emph{(\bibinfo{series}{VLDB '04})}.
\newblock


\bibitem[\protect\citeauthoryear{Bostock, Ogievetsky, and Heer}{Bostock
  et~al\mbox{.}}{2011}]%
        {Bostock11d3}
\bibfield{author}{\bibinfo{person}{Michael Bostock}, \bibinfo{person}{Vadim
  Ogievetsky}, {and} \bibinfo{person}{Jeffrey Heer}.}
  \bibinfo{year}{2011}\natexlab{}.
\newblock \showarticletitle{D3: Data-Driven Documents}.
\newblock \bibinfo{journal}{\emph{IEEE Transactions on Visualization and
  Computer Graphics}} \bibinfo{volume}{17}, \bibinfo{number}{12}
  (\bibinfo{year}{2011}), \bibinfo{pages}{2301--2309}.
\newblock
\urldef\tempurl%
\url{http://vis.stanford.edu/papers/d3}
\showURL{%
\tempurl}


\bibitem[\protect\citeauthoryear{Cheney, Chiticariu, and Tan}{Cheney
  et~al\mbox{.}}{2009}]%
        {provdb}
\bibfield{author}{\bibinfo{person}{James Cheney}, \bibinfo{person}{Laura
  Chiticariu}, {and} \bibinfo{person}{Wang~Chiew Tan}.}
  \bibinfo{year}{2009}\natexlab{}.
\newblock \showarticletitle{Provenance in databases: Why, how, and where}.
\newblock \bibinfo{journal}{\emph{{Foundations and Trends{\textregistered}~in
  Databases}}} \bibinfo{volume}{1}, \bibinfo{number}{4} (\bibinfo{year}{2009}),
  \bibinfo{pages}{379--474}.
\newblock


\bibitem[\protect\citeauthoryear{Chothia, Liagouris, McSherry, and
  Roscoe}{Chothia et~al\mbox{.}}{2016}]%
        {chothia:2016:explaining}
\bibfield{author}{\bibinfo{person}{Zaheer Chothia}, \bibinfo{person}{John
  Liagouris}, \bibinfo{person}{Frank McSherry}, {and} \bibinfo{person}{Timothy
  Roscoe}.} \bibinfo{year}{2016}\natexlab{}.
\newblock \showarticletitle{Explaining Outputs in Modern Data Analytics}.
\newblock \bibinfo{journal}{\emph{PVLDB}} \bibinfo{volume}{9},
  \bibinfo{number}{12} (\bibinfo{year}{2016}), \bibinfo{pages}{1137--1148}.
\newblock


\bibitem[\protect\citeauthoryear{Codd}{Codd}{1970}]%
        {codd}
\bibfield{author}{\bibinfo{person}{Edgar~F Codd}.}
  \bibinfo{year}{1970}\natexlab{}.
\newblock \showarticletitle{A relational model of data for large shared data
  banks}.
\newblock \bibinfo{journal}{\emph{Commun. ACM}} \bibinfo{volume}{13},
  \bibinfo{number}{6} (\bibinfo{year}{1970}), \bibinfo{pages}{377--387}.
\newblock


\bibitem[\protect\citeauthoryear{{Crossfilter}}{{Crossfilter}}{2015}]%
        {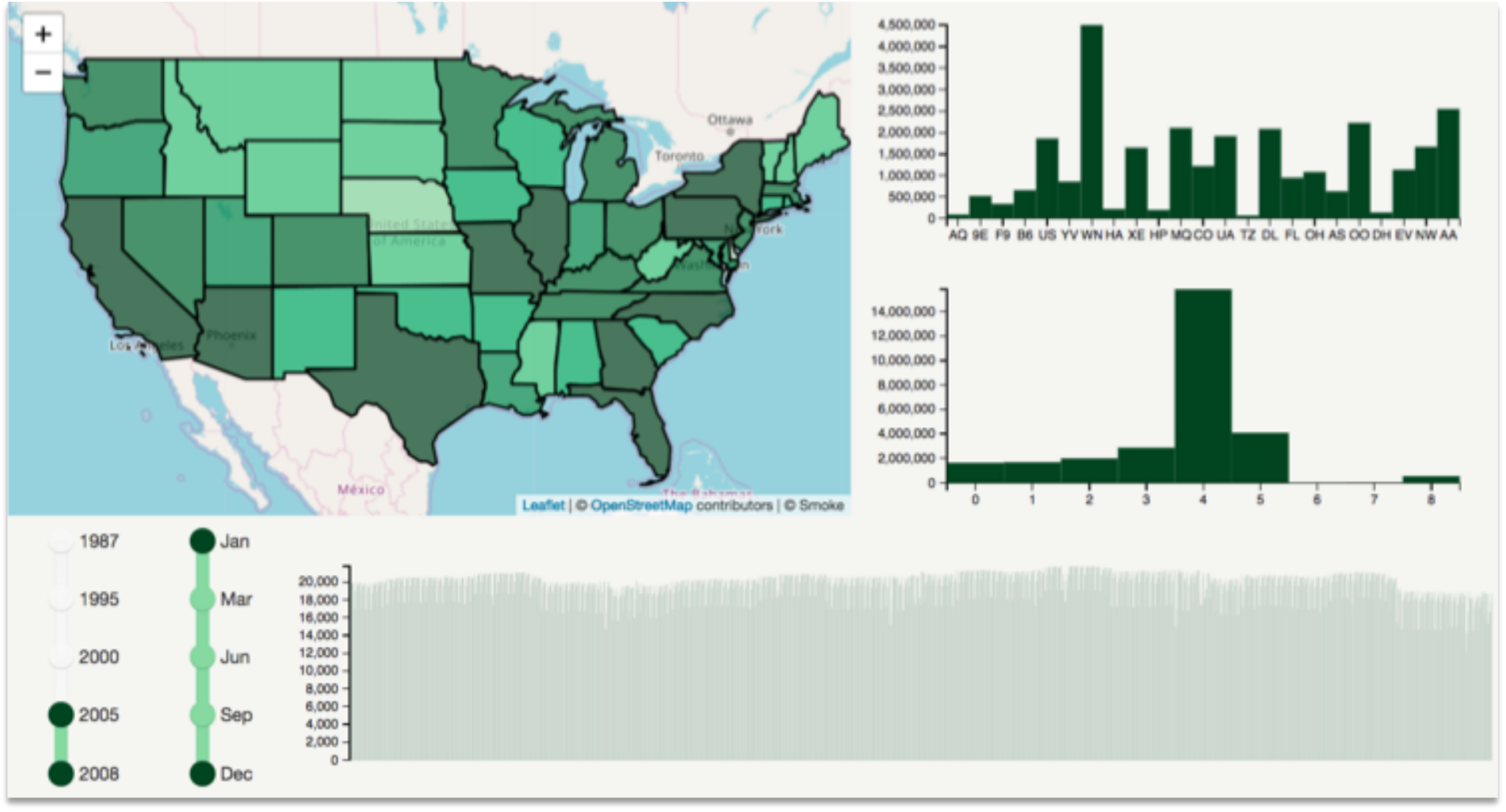}
{Crossfilter}.
\newblock \bibinfo{howpublished}{\url{http://square.github.io/crossfilter/}}.
\newblock


\bibitem[\protect\citeauthoryear{Derthick, Kolojejchick, and Roth}{Derthick
  et~al\mbox{.}}{1997}]%
        {vqe1997derthick}
\bibfield{author}{\bibinfo{person}{Mark Derthick}, \bibinfo{person}{John
  Kolojejchick}, {and} \bibinfo{person}{Steven~F. Roth}.}
  \bibinfo{year}{1997}\natexlab{}.
\newblock \showarticletitle{An Interactive Visual Query Environment for
  Exploring Data}. In \bibinfo{booktitle}{\emph{Proceedings of the 10th Annual
  ACM Symposium on User Interface Software and Technology}}
  \emph{(\bibinfo{series}{UIST '97})}.
\newblock


\bibitem[\protect\citeauthoryear{Deutch, Ives, Milo, and Tannen}{Deutch
  et~al\mbox{.}}{2013}]%
        {deutch2013caravan}
\bibfield{author}{\bibinfo{person}{Daniel Deutch}, \bibinfo{person}{Zachary~G
  Ives}, \bibinfo{person}{Tova Milo}, {and} \bibinfo{person}{Val Tannen}.}
  \bibinfo{year}{2013}\natexlab{}.
\newblock \showarticletitle{Caravan: Provisioning for What-If Analysis.}. In
  \bibinfo{booktitle}{\emph{Proceedings of the 6th biennial Conference on
  Innovative Data Systems Research}} \emph{(\bibinfo{series}{CIDR '13})}.
\newblock


\bibitem[\protect\citeauthoryear{Ebaid, Elmagarmid, Ilyas, Ouzzani,
  Quiane-Ruiz, Tang, and Yin}{Ebaid et~al\mbox{.}}{2013}]%
        {nadeef:2013:ebaid:2013}
\bibfield{author}{\bibinfo{person}{Amr Ebaid}, \bibinfo{person}{Ahmed
  Elmagarmid}, \bibinfo{person}{Ihab~F. Ilyas}, \bibinfo{person}{Mourad
  Ouzzani}, \bibinfo{person}{Jorge-Arnulfo Quiane-Ruiz}, \bibinfo{person}{Nan
  Tang}, {and} \bibinfo{person}{Si Yin}.} \bibinfo{year}{2013}\natexlab{}.
\newblock \showarticletitle{NADEEF: A Generalized Data Cleaning System}.
\newblock \bibinfo{journal}{\emph{Proceedings of the VLDB Endowment}}
  \bibinfo{volume}{6}, \bibinfo{number}{12} (\bibinfo{year}{2013}),
  \bibinfo{pages}{1218--1221}.
\newblock


\bibitem[\protect\citeauthoryear{Eugene, Leilani, and Samuel}{Eugene
  et~al\mbox{.}}{2014}]%
        {EugeneWuVision2014}
\bibfield{author}{\bibinfo{person}{Wu Eugene}, \bibinfo{person}{Battle
  Leilani}, {and} \bibinfo{person}{R.~Madden Samuel}.}
  \bibinfo{year}{2014}\natexlab{}.
\newblock \showarticletitle{The Case for Data Visualization Management
  Systems}.
\newblock \bibinfo{journal}{\emph{Proceedings of the VLDB Endowment}}
  \bibinfo{volume}{7}, \bibinfo{number}{10} (\bibinfo{year}{2014}),
  \bibinfo{pages}{903--906}.
\newblock


\bibitem[\protect\citeauthoryear{Green and Tannen}{Green and Tannen}{2017}]%
        {tannen2017pods}
\bibfield{author}{\bibinfo{person}{Todd~J. Green} {and} \bibinfo{person}{Val
  Tannen}.} \bibinfo{year}{2017}\natexlab{}.
\newblock \showarticletitle{The Semiring Franework for Database Provenance}. In
  \bibinfo{booktitle}{\emph{Proceedings of the 36th ACM SIGMOD-SIGACT-SIGAI
  Symposium on Principles of Database Systems}} \emph{(\bibinfo{series}{PODS
  '17})}.
\newblock
\urldef\tempurl%
\url{http://www.cis.upenn.edu/~val/15MayPODS.pdf}
\showURL{%
\tempurl}


\bibitem[\protect\citeauthoryear{Halevy, Korn, Noy, Olston, Polyzotis, Roy, and
  Whang}{Halevy et~al\mbox{.}}{2016}]%
        {halevy2016goods}
\bibfield{author}{\bibinfo{person}{Alon Halevy}, \bibinfo{person}{Flip Korn},
  \bibinfo{person}{Natalya~F Noy}, \bibinfo{person}{Christopher Olston},
  \bibinfo{person}{Neoklis Polyzotis}, \bibinfo{person}{Sudip Roy}, {and}
  \bibinfo{person}{Steven~Euijong Whang}.} \bibinfo{year}{2016}\natexlab{}.
\newblock \showarticletitle{Goods: Organizing google's datasets}. In
  \bibinfo{booktitle}{\emph{Proceedings of the 2016 International Conference on
  Management of Data}} \emph{(\bibinfo{series}{SIGMOD '16})}.
\newblock


\bibitem[\protect\citeauthoryear{Hanrahan}{Hanrahan}{2012}]%
        {hanrahan2012adt}
\bibfield{author}{\bibinfo{person}{Pat Hanrahan}.}
  \bibinfo{year}{2012}\natexlab{}.
\newblock \showarticletitle{Analytic Database Technologies for a New Kind of
  User: The Data Enthusiast}. In \bibinfo{booktitle}{\emph{Proceedings of the
  2012 ACM SIGMOD International Conference on Management of Data}}
  \emph{(\bibinfo{series}{SIGMOD '12})}.
\newblock


\bibitem[\protect\citeauthoryear{Harper and Agrawala}{Harper and
  Agrawala}{2014}]%
        {harper:2014:d3constructing}
\bibfield{author}{\bibinfo{person}{Jonathan Harper} {and}
  \bibinfo{person}{Maneesh Agrawala}.} \bibinfo{year}{2014}\natexlab{}.
\newblock \showarticletitle{Deconstructing and Restyling D3 Visualizations}. In
  \bibinfo{booktitle}{\emph{Proceedings of the 27th Annual ACM Symposium on
  User Interface Software and Technology}} \emph{(\bibinfo{series}{UIST '14})}.
\newblock


\bibitem[\protect\citeauthoryear{Heer, Agrawala, and Willett}{Heer
  et~al\mbox{.}}{2008}]%
        {heer2008generalized}
\bibfield{author}{\bibinfo{person}{Jeffrey Heer}, \bibinfo{person}{Maneesh
  Agrawala}, {and} \bibinfo{person}{Wesley Willett}.}
  \bibinfo{year}{2008}\natexlab{}.
\newblock \showarticletitle{Generalized selection via interactive query
  relaxation}. In \bibinfo{booktitle}{\emph{Proceedings of the SIGCHI
  Conference on Human Factors in Computing Systems}}
  \emph{(\bibinfo{series}{CHI '08})}.
\newblock


\bibitem[\protect\citeauthoryear{Heer and Shneiderman}{Heer and
  Shneiderman}{2012}]%
        {heer2012idv}
\bibfield{author}{\bibinfo{person}{Jeffrey Heer} {and} \bibinfo{person}{Ben
  Shneiderman}.} \bibinfo{year}{2012}\natexlab{}.
\newblock \showarticletitle{Interactive Dynamics for Visual Analysis}.
\newblock \bibinfo{journal}{\emph{Commun. ACM}} \bibinfo{volume}{55},
  \bibinfo{number}{4} (\bibinfo{year}{2012}), \bibinfo{pages}{45--54}.
\newblock


\bibitem[\protect\citeauthoryear{Hellerstein, Sreekanti, Gonzalez, Dalton, Dey,
  Nag, Ramachandran, Arora, Bhattacharyya, Das, et~al\mbox{.}}{Hellerstein
  et~al\mbox{.}}{2017}]%
        {hellerstein2017ground}
\bibfield{author}{\bibinfo{person}{Joseph~M Hellerstein},
  \bibinfo{person}{Vikram Sreekanti}, \bibinfo{person}{Joseph~E Gonzalez},
  \bibinfo{person}{James Dalton}, \bibinfo{person}{Akon Dey},
  \bibinfo{person}{Sreyashi Nag}, \bibinfo{person}{Krishna Ramachandran},
  \bibinfo{person}{Sudhanshu Arora}, \bibinfo{person}{Arka Bhattacharyya},
  \bibinfo{person}{Shirshanka Das}, {et~al\mbox{.}}}
  \bibinfo{year}{2017}\natexlab{}.
\newblock \showarticletitle{Ground: A Data Context Service.}. In
  \bibinfo{booktitle}{\emph{Proceedings of the 8th biennial Conference on
  Innovative Data Systems Research}} \emph{(\bibinfo{series}{CIDR '17})}.
\newblock


\bibitem[\protect\citeauthoryear{Herschel and Hlawatsch}{Herschel and
  Hlawatsch}{2016}]%
        {herschel:2016:provviz_tutorial}
\bibfield{author}{\bibinfo{person}{Melanie Herschel} {and}
  \bibinfo{person}{Marcel Hlawatsch}.} \bibinfo{year}{2016}\natexlab{}.
\newblock \showarticletitle{Provenance: On and Behind the Screens}. In
  \bibinfo{booktitle}{\emph{Proceedings of the 2016 International Conference on
  Management of Data}} \emph{(\bibinfo{series}{SIGMOD '16})}.
\newblock


\bibitem[\protect\citeauthoryear{Ikeda}{Ikeda}{2012}]%
        {ikedathesis}
\bibfield{author}{\bibinfo{person}{Robert Ikeda}.}
  \bibinfo{year}{2012}\natexlab{}.
\newblock \emph{\bibinfo{title}{Provenance In Data-Oriented Workflows}}.
\newblock \bibinfo{thesistype}{Ph.D. Dissertation}. \bibinfo{school}{Stanford
  University}.
\newblock


\bibitem[\protect\citeauthoryear{Kamat, Jayachandran, Tunga, and Nandi}{Kamat
  et~al\mbox{.}}{2014}]%
        {dice2014}
\bibfield{author}{\bibinfo{person}{Niranjan Kamat}, \bibinfo{person}{Prasanth
  Jayachandran}, \bibinfo{person}{Kathik Tunga}, {and} \bibinfo{person}{Arnab
  Nandi}.} \bibinfo{year}{2014}\natexlab{}.
\newblock \showarticletitle{Distributed and Interactive Cube Exploration}. In
  \bibinfo{booktitle}{\emph{Proceedings of the 30th International Conference on
  Data Engineering}} \emph{(\bibinfo{series}{ICDE '14})}.
\newblock


\bibitem[\protect\citeauthoryear{Kandel, Paepcke, Hellerstein, and Heer}{Kandel
  et~al\mbox{.}}{2011}]%
        {kandel2011wrangler}
\bibfield{author}{\bibinfo{person}{Sean Kandel}, \bibinfo{person}{Andreas
  Paepcke}, \bibinfo{person}{Joseph Hellerstein}, {and}
  \bibinfo{person}{Jeffrey Heer}.} \bibinfo{year}{2011}\natexlab{}.
\newblock \showarticletitle{Wrangler: interactive visual specification of data
  transformation scripts}. In \bibinfo{booktitle}{\emph{Proceedings of the
  SIGCHI Conference on Human Factors in Computing Systems}}
  \emph{(\bibinfo{series}{CHI '11})}.
\newblock


\bibitem[\protect\citeauthoryear{Kandel, Parikh, Paepcke, Hellerstein, and
  Heer}{Kandel et~al\mbox{.}}{2012}]%
        {kandel2012profiler}
\bibfield{author}{\bibinfo{person}{Sean Kandel}, \bibinfo{person}{Ravi Parikh},
  \bibinfo{person}{Andreas Paepcke}, \bibinfo{person}{Joseph~M Hellerstein},
  {and} \bibinfo{person}{Jeffrey Heer}.} \bibinfo{year}{2012}\natexlab{}.
\newblock \showarticletitle{Profiler: Integrated statistical analysis and
  visualization for data quality assessment}. In
  \bibinfo{booktitle}{\emph{Proceedings of the International Working Conference
  on Advanced Visual Interfaces}} \emph{(\bibinfo{series}{AVI '12})}.
\newblock


\bibitem[\protect\citeauthoryear{Kim, Blais, Parameswaran, Indyk, Madden, and
  Rubinfeld}{Kim et~al\mbox{.}}{2014}]%
        {kim2014rapid}
\bibfield{author}{\bibinfo{person}{Albert Kim}, \bibinfo{person}{Eric Blais},
  \bibinfo{person}{Aditya Parameswaran}, \bibinfo{person}{Piotr Indyk},
  \bibinfo{person}{Sam Madden}, {and} \bibinfo{person}{Ronitt Rubinfeld}.}
  \bibinfo{year}{2014}\natexlab{}.
\newblock \showarticletitle{Rapid Sampling for Visualizations with Ordering
  Guarantees}.
\newblock \bibinfo{journal}{\emph{Proceedings of the VLDB Endowment}}
  \bibinfo{volume}{8}, \bibinfo{number}{5} (\bibinfo{year}{2014}),
  \bibinfo{pages}{521--532}.
\newblock


\bibitem[\protect\citeauthoryear{Lee, Kohler, Ludascher, and Glavic}{Lee
  et~al\mbox{.}}{2017}]%
        {gpromwhynot2017lee}
\bibfield{author}{\bibinfo{person}{S. Lee}, \bibinfo{person}{S. Kohler},
  \bibinfo{person}{B. Ludascher}, {and} \bibinfo{person}{B. Glavic}.}
  \bibinfo{year}{2017}\natexlab{}.
\newblock \showarticletitle{A SQL-Middleware Unifying Why and Why-Not
  Provenance for First-Order Queries}. In \bibinfo{booktitle}{\emph{Proceedings
  of the 33rd International Conference on Data Engineering}}
  \emph{(\bibinfo{series}{ICDE '17})}.
\newblock


\bibitem[\protect\citeauthoryear{Lins, Klosowski, and Scheidegger}{Lins
  et~al\mbox{.}}{2013}]%
        {lins2013nanocubes}
\bibfield{author}{\bibinfo{person}{Lauro Lins}, \bibinfo{person}{James~T
  Klosowski}, {and} \bibinfo{person}{Carlos Scheidegger}.}
  \bibinfo{year}{2013}\natexlab{}.
\newblock \showarticletitle{Nanocubes for Real-Time Exploration of
  Spatiotemporal Datasets}.
\newblock \bibinfo{journal}{\emph{EuroVis}} \bibinfo{volume}{19},
  \bibinfo{number}{12} (\bibinfo{year}{2013}), \bibinfo{pages}{2456--2465}.
\newblock


\bibitem[\protect\citeauthoryear{Liu, Jiang, and Heer}{Liu
  et~al\mbox{.}}{2013}]%
        {liu2013immens}
\bibfield{author}{\bibinfo{person}{Zhicheng Liu}, \bibinfo{person}{Biye Jiang},
  {and} \bibinfo{person}{Jeffrey Heer}.} \bibinfo{year}{2013}\natexlab{}.
\newblock \showarticletitle{{imMens}: Real-time Visual Querying of Big Data}.
\newblock \bibinfo{journal}{\emph{Computer Graphics Forum}}
  \bibinfo{volume}{32}, \bibinfo{number}{3pt4} (\bibinfo{year}{2013}),
  \bibinfo{pages}{421--430}.
\newblock


\bibitem[\protect\citeauthoryear{Livny, Ramakrishnan, Beyer, Chen, Donjerkovic,
  Lawande, Myllymaki, and Wenger}{Livny et~al\mbox{.}}{1997}]%
        {livny1997devise}
\bibfield{author}{\bibinfo{person}{M Livny}, \bibinfo{person}{R Ramakrishnan},
  \bibinfo{person}{K Beyer}, \bibinfo{person}{G Chen}, \bibinfo{person}{D
  Donjerkovic}, \bibinfo{person}{S Lawande}, \bibinfo{person}{J Myllymaki},
  {and} \bibinfo{person}{K Wenger}.} \bibinfo{year}{1997}\natexlab{}.
\newblock \showarticletitle{DEVise: Integrated Querying and Visual Exploration
  of Large Datasets (Demo Abstract)}. In \bibinfo{booktitle}{\emph{Proceedings
  of the 1997 ACM SIGMOD International Conference on Management of Data}}
  \emph{(\bibinfo{series}{SIGMOD '97})}.
\newblock


\bibitem[\protect\citeauthoryear{Mavlyutov, Curino, Asipov, and
  Cudre-Mauroux}{Mavlyutov et~al\mbox{.}}{2017}]%
        {mavlyutov2017dependency}
\bibfield{author}{\bibinfo{person}{Ruslan Mavlyutov}, \bibinfo{person}{Carlo
  Curino}, \bibinfo{person}{Boris Asipov}, {and} \bibinfo{person}{Philippe
  Cudre-Mauroux}.} \bibinfo{year}{2017}\natexlab{}.
\newblock \showarticletitle{Dependency-Driven Analytics: A Compass for
  Uncharted Data Oceans.}. In \bibinfo{booktitle}{\emph{Proceedings of the 8th
  biennial Conference on Innovative Data Systems Research}}
  \emph{(\bibinfo{series}{CIDR '17})}.
\newblock


\bibitem[\protect\citeauthoryear{Mottin, Lissandrini, Velegrakis, and
  Palpanas}{Mottin et~al\mbox{.}}{2014}]%
        {mottin2014eqg}
\bibfield{author}{\bibinfo{person}{Davide Mottin}, \bibinfo{person}{Matteo
  Lissandrini}, \bibinfo{person}{Yannis Velegrakis}, {and}
  \bibinfo{person}{Themis Palpanas}.} \bibinfo{year}{2014}\natexlab{}.
\newblock \showarticletitle{Exemplar Queries: Give Me an Example of What You
  Need}.
\newblock \bibinfo{journal}{\emph{Proceedings of the VLDB Endowment}}
  \bibinfo{volume}{7}, \bibinfo{number}{5} (\bibinfo{year}{2014}),
  \bibinfo{pages}{365--376}.
\newblock


\bibitem[\protect\citeauthoryear{Niu, Kapoor, Glavic, Gawlick, Liu,
  Krishnaswamy, and Radhakrishnan}{Niu et~al\mbox{.}}{2017}]%
        {gprom}
\bibfield{author}{\bibinfo{person}{Xing Niu}, \bibinfo{person}{Raghav Kapoor},
  \bibinfo{person}{Boris Glavic}, \bibinfo{person}{Dieter Gawlick},
  \bibinfo{person}{Zhen~Hua Liu}, \bibinfo{person}{Vasudha Krishnaswamy}, {and}
  \bibinfo{person}{Venkatesh Radhakrishnan}.} \bibinfo{year}{2017}\natexlab{}.
\newblock \showarticletitle{Provenance-aware Query Optimization}. In
  \bibinfo{booktitle}{\emph{Proceedings of the 33rd International Conference on
  Data Engineering}} \emph{(\bibinfo{series}{ICDE '17})}.
  \bibinfo{pages}{473--484}.
\newblock


\bibitem[\protect\citeauthoryear{North and Shneiderman}{North and
  Shneiderman}{2000}]%
        {north2000snap}
\bibfield{author}{\bibinfo{person}{Chris North} {and} \bibinfo{person}{Ben
  Shneiderman}.} \bibinfo{year}{2000}\natexlab{}.
\newblock \showarticletitle{Snap-together visualization: a user interface for
  coordinating visualizations via relational schemata}. In
  \bibinfo{booktitle}{\emph{Proceedings of the Working Conference on Advanced
  Visual Interfaces}} \emph{(\bibinfo{series}{AVI '00})}.
\newblock


\bibitem[\protect\citeauthoryear{Ontime}{Ontime}{[n. d.]}]%
        {ontime}
Ontime.
\newblock
  \bibinfo{howpublished}{\url{http://stat-computing.org/dataexpo/2009/the-data.html}}.
\newblock


\bibitem[\protect\citeauthoryear{Oracle}{Oracle}{2014}]%
        {endeca}
\bibfield{author}{\bibinfo{person}{Oracle}.} \bibinfo{year}{2014}\natexlab{}.
\newblock \bibinfo{booktitle}{\emph{Oracle Endeca Information Discovery: A
  Technical Overview}}.
\newblock \bibinfo{type}{{T}echnical {R}eport}. \bibinfo{institution}{Oracle}.
\newblock


\bibitem[\protect\citeauthoryear{Papenbrock, Bergmann, Finke, Zwiener, and
  Naumann}{Papenbrock et~al\mbox{.}}{2015}]%
        {metanome:2015:papenbrock}
\bibfield{author}{\bibinfo{person}{Thorsten Papenbrock}, \bibinfo{person}{Tanja
  Bergmann}, \bibinfo{person}{Moritz Finke}, \bibinfo{person}{Jakob Zwiener},
  {and} \bibinfo{person}{Felix Naumann}.} \bibinfo{year}{2015}\natexlab{}.
\newblock \showarticletitle{Data Profiling with Metanome}.
\newblock \bibinfo{journal}{\emph{Proceedings of the VLDB Endowment}}
  \bibinfo{volume}{8}, \bibinfo{number}{12} (\bibinfo{year}{2015}),
  \bibinfo{pages}{1860--1863}.
\newblock


\bibitem[\protect\citeauthoryear{Power BI}{Power BI}{2018}]%
        {powerbi}
Power BI.
\newblock \bibinfo{howpublished}{\url{https://powerbi.microsoft.com}}.
\newblock


\bibitem[\protect\citeauthoryear{Procopio, Scheidegger, Wu, and Chang}{Procopio
  et~al\mbox{.}}{2017}]%
        {procopioload}
\bibfield{author}{\bibinfo{person}{Marianne Procopio}, \bibinfo{person}{Carlos
  Scheidegger}, \bibinfo{person}{Eugene Wu}, {and} \bibinfo{person}{Remco
  Chang}.} \bibinfo{year}{2017}\natexlab{}.
\newblock \showarticletitle{Load-n-Go: Fast Approximate Join Visualizations
  That Improve Over Time}. In \bibinfo{booktitle}{\emph{Proceedings of the 2nd
  Workshop on Data Systems for Interactive Analysis}}
  \emph{(\bibinfo{series}{DSIA '17})}.
\newblock


\bibitem[\protect\citeauthoryear{Psallidas, Ding, Chakrabarti, and
  Chaudhuri}{Psallidas et~al\mbox{.}}{2015}]%
        {psallidas2015s4}
\bibfield{author}{\bibinfo{person}{Fotis Psallidas}, \bibinfo{person}{Bolin
  Ding}, \bibinfo{person}{Kaushik Chakrabarti}, {and} \bibinfo{person}{Surajit
  Chaudhuri}.} \bibinfo{year}{2015}\natexlab{}.
\newblock \showarticletitle{S4: Top-k Spreadsheet-Style Search for Query
  Discovery}. In \bibinfo{booktitle}{\emph{Proceedings of the 2015 ACM SIGMOD
  International Conference on Management of Data}}
  \emph{(\bibinfo{series}{SIGMOD '15})}.
\newblock


\bibitem[\protect\citeauthoryear{Psallidas and Wu}{Psallidas and Wu}{2018a}]%
        {psallidas2018smokedemo}
\bibfield{author}{\bibinfo{person}{Fotis Psallidas} {and}
  \bibinfo{person}{Eugene Wu}.} \bibinfo{year}{2018}\natexlab{a}.
\newblock \showarticletitle{Demonstration of Smoke: A Deep Breath of
  Data-Intensive Lineage Applications}. In
  \bibinfo{booktitle}{\emph{Proceedings of the 2018 ACM SIGMOD International
  Conference on Management of Data}} \emph{(\bibinfo{series}{SIGMOD '18})}.
\newblock


\bibitem[\protect\citeauthoryear{Psallidas and Wu}{Psallidas and Wu}{2018b}]%
        {psallidas2018smoke}
\bibfield{author}{\bibinfo{person}{Fotis Psallidas} {and}
  \bibinfo{person}{Eugene Wu}.} \bibinfo{year}{2018}\natexlab{b}.
\newblock \showarticletitle{Smoke: Fined-Grained Lineage Capture At Interactive
  Speed}.
\newblock \bibinfo{journal}{\emph{Proceedings of the VLDB Endowment}}
  \bibinfo{volume}{11}, \bibinfo{number}{6} (\bibinfo{year}{2018}),
  \bibinfo{pages}{719--732}.
\newblock


\bibitem[\protect\citeauthoryear{Ragan, Endert, Sanyal, and Chen}{Ragan
  et~al\mbox{.}}{2016}]%
        {ragan2016characterizing}
\bibfield{author}{\bibinfo{person}{Eric~D Ragan}, \bibinfo{person}{Alex
  Endert}, \bibinfo{person}{Jibonananda Sanyal}, {and} \bibinfo{person}{Jian
  Chen}.} \bibinfo{year}{2016}\natexlab{}.
\newblock \showarticletitle{Characterizing provenance in visualization and data
  analysis: an organizational framework of provenance types and purposes}.
\newblock \bibinfo{journal}{\emph{IEEE Transactions on Visualization and
  Computer Graphics}} \bibinfo{volume}{22}, \bibinfo{number}{1}
  (\bibinfo{year}{2016}), \bibinfo{pages}{31--40}.
\newblock


\bibitem[\protect\citeauthoryear{Rahman, Aliakbarpour, Kong, Blais, Karahalios,
  Parameswaran, and Rubinfield}{Rahman et~al\mbox{.}}{2017}]%
        {rahman2017ve}
\bibfield{author}{\bibinfo{person}{Sajjadur Rahman}, \bibinfo{person}{Maryam
  Aliakbarpour}, \bibinfo{person}{Ha~Kyung Kong}, \bibinfo{person}{Eric Blais},
  \bibinfo{person}{Karrie Karahalios}, \bibinfo{person}{Aditya Parameswaran},
  {and} \bibinfo{person}{Ronitt Rubinfield}.} \bibinfo{year}{2017}\natexlab{}.
\newblock \showarticletitle{I've seen enough: incrementally improving
  visualizations to support rapid decision making}.
\newblock \bibinfo{journal}{\emph{Proceedings of the VLDB Endowment}}
  \bibinfo{volume}{10}, \bibinfo{number}{11} (\bibinfo{year}{2017}),
  \bibinfo{pages}{1262--1273}.
\newblock


\bibitem[\protect\citeauthoryear{Roy, Orr, and Suciu}{Roy
  et~al\mbox{.}}{2015}]%
        {roy2015explain}
\bibfield{author}{\bibinfo{person}{Sudeepa Roy}, \bibinfo{person}{Laurel Orr},
  {and} \bibinfo{person}{Dan Suciu}.} \bibinfo{year}{2015}\natexlab{}.
\newblock \showarticletitle{Explaining Query Answers with Explanation-ready
  Databases}.
\newblock \bibinfo{journal}{\emph{Proceedings of the VLDB Endowment}}
  \bibinfo{volume}{9}, \bibinfo{number}{4} (\bibinfo{year}{2015}),
  \bibinfo{pages}{348--359}.
\newblock


\bibitem[\protect\citeauthoryear{RStudio Shiny}{RStudio Shiny}{2016}]%
        {shiny}
RStudio Shiny.
\newblock \bibinfo{title}{RStudio Shiny}.
\newblock \bibinfo{howpublished}{https://shiny.rstudio.com/}.
\newblock


\bibitem[\protect\citeauthoryear{Satyanarayan, Moritz, Wongsuphasawat, and
  Heer}{Satyanarayan et~al\mbox{.}}{2017}]%
        {satyanarayan2017vegalite}
\bibfield{author}{\bibinfo{person}{Arvind Satyanarayan},
  \bibinfo{person}{Dominik Moritz}, \bibinfo{person}{Kanit Wongsuphasawat},
  {and} \bibinfo{person}{Jeffrey Heer}.} \bibinfo{year}{2017}\natexlab{}.
\newblock \showarticletitle{Vega-lite: A grammar of interactive graphics}.
\newblock \bibinfo{journal}{\emph{Transactions on Visualization and Computer
  Graphics}} \bibinfo{volume}{23}, \bibinfo{number}{1} (\bibinfo{year}{2017}),
  \bibinfo{pages}{341--350}.
\newblock


\bibitem[\protect\citeauthoryear{Satyanarayan, Russell, Hoffswell, and
  Heer}{Satyanarayan et~al\mbox{.}}{2016}]%
        {reactivevega}
\bibfield{author}{\bibinfo{person}{Arvind Satyanarayan}, \bibinfo{person}{Ryan
  Russell}, \bibinfo{person}{Jane Hoffswell}, {and} \bibinfo{person}{Jeffrey
  Heer}.} \bibinfo{year}{2016}\natexlab{}.
\newblock \showarticletitle{Reactive Vega: A Streaming Dataflow Architecture
  for Declarative Interactive Visualization}.
\newblock \bibinfo{journal}{\emph{Transactions on Visualization and Computer
  Graphics (Proc. InfoVis)}} (\bibinfo{year}{2016}).
\newblock


\bibitem[\protect\citeauthoryear{Scheidegger, Vo, Koop, Freire, and
  Silva}{Scheidegger et~al\mbox{.}}{2007}]%
        {scheidegger2007querying}
\bibfield{author}{\bibinfo{person}{Carlos Scheidegger}, \bibinfo{person}{Huy
  Vo}, \bibinfo{person}{David Koop}, \bibinfo{person}{Juliana Freire}, {and}
  \bibinfo{person}{Claudio Silva}.} \bibinfo{year}{2007}\natexlab{}.
\newblock \showarticletitle{Querying and creating visualizations by analogy}.
\newblock \bibinfo{journal}{\emph{Transactions on Visualization and Computer
  Graphics}} \bibinfo{volume}{13}, \bibinfo{number}{6} (\bibinfo{year}{2007}),
  \bibinfo{pages}{1560--1567}.
\newblock


\bibitem[\protect\citeauthoryear{Shneiderman}{Shneiderman}{1984}]%
        {shneiderman:1984}
\bibfield{author}{\bibinfo{person}{Ben Shneiderman}.}
  \bibinfo{year}{1984}\natexlab{}.
\newblock \showarticletitle{Response Time and Display Rate in Human Performance
  with Computers}.
\newblock \bibinfo{journal}{\emph{CSUR}} (\bibinfo{year}{1984}).
\newblock


\bibitem[\protect\citeauthoryear{{Strobelt}, {Gehrmann}, {Behrisch}, {Perer},
  {Pfister}, and {Rush}}{{Strobelt} et~al\mbox{.}}{2018}]%
        {seq2seqvisv1}
\bibfield{author}{\bibinfo{person}{H. {Strobelt}}, \bibinfo{person}{S.
  {Gehrmann}}, \bibinfo{person}{M. {Behrisch}}, \bibinfo{person}{A. {Perer}},
  \bibinfo{person}{H. {Pfister}}, {and} \bibinfo{person}{A.~M. {Rush}}.}
  \bibinfo{year}{2018}\natexlab{}.
\newblock \showarticletitle{{Seq2Seq-Vis: A Visual Debugging Tool for
  Sequence-to-Sequence Models}}.
\newblock \bibinfo{journal}{\emph{ArXiv e-prints}} (\bibinfo{year}{2018}).
\newblock
\showeprint[arxiv]{1804.09299v1}


\bibitem[\protect\citeauthoryear{Tensorboard: Visualizing
  Learning}{Tensorboard: Visualizing Learning}{2016}]%
        {tensorboard}
Tensorboard: Visualizing Learning.
\newblock \bibinfo{howpublished}{\url{http://tinyurl.com/zayyabk}}.
\newblock


\bibitem[\protect\citeauthoryear{Tukey}{Tukey}{1977}]%
        {tukey1977exploratory}
\bibfield{author}{\bibinfo{person}{John~W Tukey}.}
  \bibinfo{year}{1977}\natexlab{}.
\newblock \bibinfo{booktitle}{\emph{Exploratory data analysis}}.
\newblock \bibinfo{publisher}{Reading, Mass.}
\newblock


\bibitem[\protect\citeauthoryear{Tylenda, Sozio, and Weikum}{Tylenda
  et~al\mbox{.}}{2011}]%
        {yagobrowswer2011tylenda}
\bibfield{author}{\bibinfo{person}{Tomasz Tylenda}, \bibinfo{person}{Mauro
  Sozio}, {and} \bibinfo{person}{Gerhard Weikum}.}
  \bibinfo{year}{2011}\natexlab{}.
\newblock \showarticletitle{Einstein: Physicist or Vegetarian? Summarizing
  Semantic Type Graphs for Knowledge Discovery}. In
  \bibinfo{booktitle}{\emph{Proceedings of the 20th International Conference
  Companion on World Wide Web}} \emph{(\bibinfo{series}{WWW '11})}.
\newblock


\bibitem[\protect\citeauthoryear{Wilhelm}{Wilhelm}{2003}]%
        {wilhelm2003taxonomy}
\bibfield{author}{\bibinfo{person}{Adalbert Wilhelm}.}
  \bibinfo{year}{2003}\natexlab{}.
\newblock \showarticletitle{User Interaction at Various Levels of Data
  Displays}. In \bibinfo{booktitle}{\emph{CSDA}}.
\newblock


\bibitem[\protect\citeauthoryear{Woodruff and Stonebraker}{Woodruff and
  Stonebraker}{1997}]%
        {woodruff1997supporting}
\bibfield{author}{\bibinfo{person}{Allison Woodruff} {and}
  \bibinfo{person}{Michael Stonebraker}.} \bibinfo{year}{1997}\natexlab{}.
\newblock \showarticletitle{Supporting Fine-grained Data Lineage in a Database
  Visualization Environment}. In \bibinfo{booktitle}{\emph{Proceedings of the
  13th International Conference on Data Engineering}}
  \emph{(\bibinfo{series}{ICDE '97})}.
\newblock


\bibitem[\protect\citeauthoryear{Wu and Madden}{Wu and Madden}{2013}]%
        {scorpion}
\bibfield{author}{\bibinfo{person}{Eugene Wu} {and} \bibinfo{person}{Samuel
  Madden}.} \bibinfo{year}{2013}\natexlab{}.
\newblock \showarticletitle{Scorpion: Explaining Away Outliers in Aggregate
  Queries}.
\newblock \bibinfo{journal}{\emph{Proceedings of the VLDB Endowment}}
  \bibinfo{volume}{6}, \bibinfo{number}{8} (\bibinfo{year}{2013}),
  \bibinfo{pages}{553--564}.
\newblock


\bibitem[\protect\citeauthoryear{Wu, Madden, and Stonebraker}{Wu
  et~al\mbox{.}}{2012}]%
        {wu2012demonstration}
\bibfield{author}{\bibinfo{person}{Eugene Wu}, \bibinfo{person}{Samuel Madden},
  {and} \bibinfo{person}{Michael Stonebraker}.}
  \bibinfo{year}{2012}\natexlab{}.
\newblock \showarticletitle{A demonstration of DBWipes: clean as you query}.
\newblock \bibinfo{journal}{\emph{Proceedings of the VLDB Endowment}}
  \bibinfo{volume}{5}, \bibinfo{number}{12} (\bibinfo{year}{2012}),
  \bibinfo{pages}{1894--1897}.
\newblock


\bibitem[\protect\citeauthoryear{Wu, Madden, and Stonebraker}{Wu
  et~al\mbox{.}}{2013}]%
        {wu2013subzero}
\bibfield{author}{\bibinfo{person}{Eugene Wu}, \bibinfo{person}{Samuel Madden},
  {and} \bibinfo{person}{Michael Stonebraker}.}
  \bibinfo{year}{2013}\natexlab{}.
\newblock \showarticletitle{Subzero: a fine-grained lineage system for
  scientific databases}. In \bibinfo{booktitle}{\emph{Proceedings of the 29th
  International Conference on Data Engineering}} \emph{(\bibinfo{series}{ICDE
  '13})}. \bibinfo{pages}{865--876}.
\newblock


\bibitem[\protect\citeauthoryear{Wu, Psallidas, Miao, Zhang, Rettig, Wu, and
  Sellam}{Wu et~al\mbox{.}}{2017}]%
        {wu2017dvms}
\bibfield{author}{\bibinfo{person}{Eugene Wu}, \bibinfo{person}{Fotis
  Psallidas}, \bibinfo{person}{Zhengjie Miao}, \bibinfo{person}{Haoci Zhang},
  \bibinfo{person}{Laura Rettig}, \bibinfo{person}{Yifan Wu}, {and}
  \bibinfo{person}{Thibault Sellam}.} \bibinfo{year}{2017}\natexlab{}.
\newblock \showarticletitle{Combining Design and Performance in a Data
  Visualization Management System}. In \bibinfo{booktitle}{\emph{Proceedings of
  the 8th biennial Conference on Innovative Data Systems Research}}
  \emph{(\bibinfo{series}{CIDR '17})}.
\newblock


\bibitem[\protect\citeauthoryear{Yi, Kang, Stasko, and Jacko}{Yi
  et~al\mbox{.}}{2007}]%
        {yi2007taxonomy}
\bibfield{author}{\bibinfo{person}{Ji~Soo Yi}, \bibinfo{person}{Youn~ah Kang},
  \bibinfo{person}{John Stasko}, {and} \bibinfo{person}{Julie Jacko}.}
  \bibinfo{year}{2007}\natexlab{}.
\newblock \showarticletitle{Toward a Deeper Understanding of the Role of
  Interaction in Information Visualization}. In
  \bibinfo{booktitle}{\emph{TVCG}}.
\newblock


\end{thebibliography}

\end{document}